%% file: main.tex
\documentclass[twocolumn]{aastex631}

\usepackage{amsmath}
\usepackage{xcolor}
\usepackage{graphicx}
\usepackage{booktabs}
\usepackage{multirow}
\usepackage{placeins}

\usepackage{hyperref}
\usepackage{layouts}
\usepackage{soul}
\usepackage{fancyvrb}

\input{macros/units.tex}

\begin{document}

\title{3D hydrodynamic simulations of massive main-sequence stars - IV. Internal gravity waves matter for SLF variability.}

\author[0009-0005-9976-1882]{Praneet Pathak}
\author[0000-0002-9632-1436]{Simon Blouin}
\author[0000-0001-8087-9278]{Falk Herwig}
\affiliation{Department of Physics $\&$ Astronomy, University of Victoria, Victoria, BC, V8W 2Y2, Canada}
\author{Paul R. Woodward}
\affiliation{LCSE and Department of Astronomy, University of Minnesota, Minneapolis, MN 55455, USA}

\begin{abstract}
The power spectrum of light curves from satellites like CoRoT and TESS of massive
main-sequence stars show stochastic low-frequency (SLF) variability. To investigate the origin of this phenomenon, we conducted
high-resolution 3D hydrodynamic \texttt{PPMstar} simulations of a non-rotating \unit{25}{\Msun} zero-age main sequence star, modeling 95\% of the stellar structure with
both a core and a thin outer envelope convection zone. The outer envelope
convection zone was implemented through modification of the opacity model, shifting the Fe
opacity bump inward and enhancing its amplitude for computational feasibility. The
luminosity power spectrum from our primary simulation (M424) exhibits qualitative and
quantitative characteristics similar to observed SLF variability, with a
 $\approx2$-dex difference between high- and low-frequency power. The spectrum displays distinct features attributable to internal gravity
wave (IGW) eigenmodes. To isolate the contributions of different stellar regions, we
performed numerical experiments with suppressed core convection, envelope
convection and envelope-only configurations. The comparative analysis demonstrates that
outer envelope convection alone produces significantly less low-frequency power than
the full-star configuration. In our simulations the outer envelope
convection zone excites at its inner boundary a rich IGW eigenmode spectrum in the layer
just below. In an otherwise identical simulation where the core convection is not driven by heating, the SLF spectrum is remarkably similar and the integrated power is reduced by only 10\%, suggesting that the envelope convection is the dominant contributor to SLF power spectrum. The IGW spectral characteristics
depend on the complete stellar stratification, demonstrating that interior structure could
influence observable surface variability.
\end{abstract}

\section{Introduction}
\label{sec:introduction}

Stochastic low-frequency (SLF) variability, characterized by excess power at lower frequencies in the power spectrum of observed light curves, is a ubiquitous phenomenon detected in the luminosity power spectra of O and B type stars \citep{Blomme2011,Bowman2019a,Bowman2019b}. This variability is observed across diverse metallicity environments, indicating a common underlying physical mechanism \citep{Bowman2024}.

Understanding this mechanism is important, as internal gravity waves (IGWs) generated in the stellar interior and propagating to the surface represent a potential source of this phenomenon. Confirmation of this hypothesis may potentially facilitate inference of internal stellar structure parameters from external light curve observations, such as convective core mass, radii, internal rotation rate, and other structural properties \citep{Aerts2021,Mombarg2024}. The implications extend to the constraint of internal stellar stratification parameters, including convective core size and composition, which would refine stellar evolution models and advance stellar physics \citep{Aerts2010}. For example, \citet{Burssens2023} used asteroseismology to deduce the convective core mass and showed non-rigid radial rotation in HD 192575, a 12 $\mathrm{M_\odot}$ main-sequence star. \citet{Pedersen2018} showed that one could differentiate between step and exponential overshooting near the convective core boundary from gravity-mode period spacings in slowly pulsating B-type stars. \citet{Papics2017} deduced internal rotational rates of five slow pulsating B stars.

The physical mechanism responsible for SLF variability is still under debate \citep[for a review see][]{Bowman2023}. Multiple 2D and 3D simulations of massive stars have reproduced frequency spectra from mock luminosity observations with morphology similar to SLF variability \citep[e.g.][]{Rogers2013,Rogers2015,Aerts2015,Rogers2017,Edelmann2019,Ratnasingam2019,Ratnasingam2020,Ratnasingam2023,Horst2020,Varghese2023,Vanon2023,Thompson2024}. Among the proposed mechanisms, core convection has been suggested as a source of gravity waves that propagate to the surface, where temperature fluctuations from the superposition of these IGWs produce luminosity variations manifesting as SLF variability in luminosity time series \citep{Aerts2015}.

However, \citet{Anders2023} conducted 3D simulations of massive star convection using a two-component approach: wave generation simulations that directly model core convection and wave excitation, combined with theoretical transfer functions to represent wave propagation through the stellar envelope to the surface. Their wave propagation simulations extended to 93\% of the stellar radius with mode lifetimes of $\lesssim 10$ years. Using this methodology, they concluded that gravity waves excited by turbulent core convection do not reach the stellar surface with observable amplitudes due to radiative damping \citep{Lecoanet2019,Lecoanet2021,LeSaux2023}.

Some investigations propose that if SLF variability does not originate from gravity waves excited by core convection, it may instead result from turbulence induced by subsurface convection due to the Fe opacity bump near the surface \citep{Cantiello2009,Cantiello2021,Schultz2022,Schultz2023}. Nevertheless, \citet{Jermyn2022} found that main-sequence stars (8-20 $\mathrm{M_\odot}$) at SMC-like metallicities lack substantial subsurface convection zones because their Rayleigh numbers fall below the critical value required for convection onset, despite being predicted as convectively unstable by 1D stellar evolution models, yet these stars exhibit SLF variability comparable to stars with such convection zones \citep{Bowman2024}. 
Stochastic light variations from wind instability constitute another potential mechanism for SLF variability in massive stars \citep{Krticka2018,Krticka2021}. This mechanism, however, is expected to have minimal impact for late O-type and early B-type main-sequence stars due to their optically thin and weak winds, particularly at low metallicity.

In our series of papers on massive main sequence stars, \citet{Thompson2024} conducted 3D hydrodynamic simulations of a 25 $\mathrm{M_\odot}$ mid-main-sequence star extending from the stellar center to approximately 54\% of the stellar radius. These simulations, performed without radiative effects, reproduced qualitatively similar spectra to observed SLF variability assuming IGWs propagate to the photosphere, demonstrating that core convection can stochastically excite IGWs that produce qualitatively SLF like spectra. However, the characteristic frequency obtained ($\mathrm{\nu_{char} \approx 6\, \mu Hz}$) was significantly smaller than typical observational values \citep{Bowman2019a,Bowman2019b,Bowman2020}. Additionally, those synthetic observations were extracted from deep within the stellar interior, far from the observable photosphere, and the simulations did not account for thin outer envelope convection zones that may contribute to surface variability. These limitations motivated the present study to investigate the combined effects of both core and thin outer envelope convection in a more complete stellar model.

In this paper, we present high-resolution \texttt{PPMstar} simulations of a non-rotating 25 $\mathrm{M_\odot}$ ZAMS (Zero Age Main Sequence) star to investigate the origin of low-frequency excess. In Section \ref{sec:Methods}, we describe our base state and the modified opacity model employed in the simulations. In Section \ref{sec:Results}, we present power spectra from simulated light curves from different runs and analyze them comprehensively. Finally, in Section \ref{sec:discussion-and-conclusion}, we discuss our findings and present conclusions outlining directions for future research.

\section{Methods}
\label{sec:Methods}

\subsection{Base \texttt{MESA} state}

This work employs the 3D hydrodynamics \texttt{PPMstar} explicit gas dynamics code, which incorporates an ideal gas plus radiation pressure equation of state with radiative diffusion in the energy flux \citep{Woodward1984,Colella1984,Woodward2015,Herwig2023,Mao2024}. Initial conditions\footnote{Please refer to the MESA profile available in the Zenodo repository \href{https://doi.org/10.5281/zenodo.15679630}{10.5281/zenodo.15679630}.} for the simulations are derived from a Zero Age Main Sequence (ZAMS) state constructed using Modules for Experiments in Stellar Astrophysics (\texttt{MESA}) revision 5329 \citep{Paxton2011}. 
\begin{figure*}
   \centering
   \includegraphics[width=0.49\hsize]{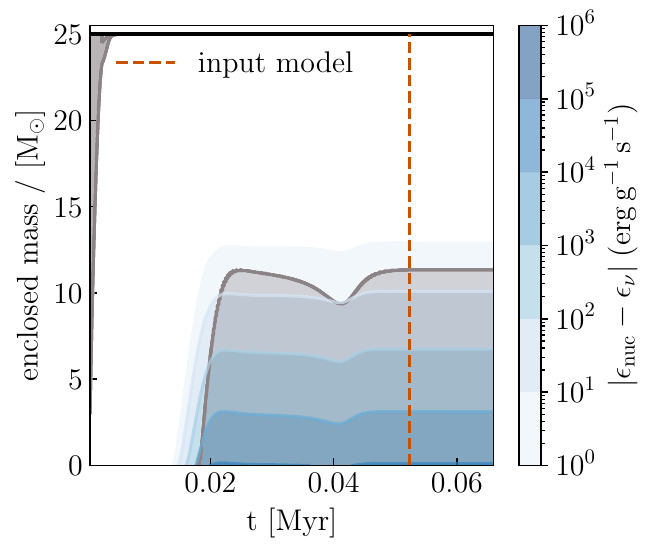}
   \includegraphics[width=0.49\hsize]{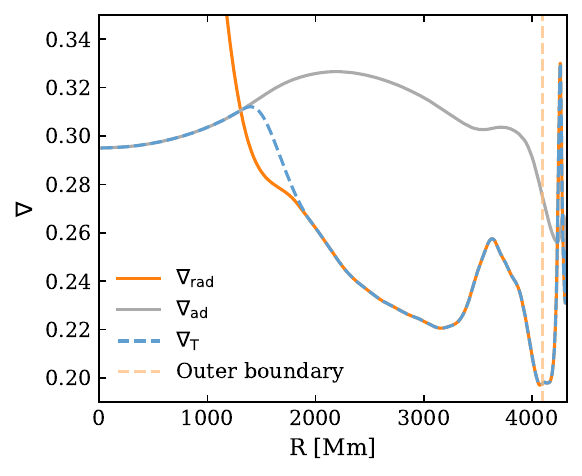}
      \caption{(\textit{Left}): Kippenhahn diagram of the ZAMS 25 $\mathrm{M_\odot}$ \texttt{MESA} model. Grey regions represent convective zones, white regions represent radiative zones, and blue contours indicate nuclear burning regions with intensities shown in the colorbar. The red dashed vertical line marks the initial model used for all the \texttt{PPMstar} simulations presented in this work. (\textit{Right}): Radial profiles of the radiative ($\nabla_{\rm rad}$, solid orange line), adiabatic ($\nabla_{\rm ad}$, grey solid line), and actual temperature gradients ($\nabla_{\rm T}$, blue dashed line) for the ZAMS \texttt{MESA} model. The light orange dashed vertical line marks the outer boundary of the simulations presented in this work.}
         \label{Fig:Kippenhahn}
\end{figure*}
The base model comprises a non-rotating 25 $\mathrm{M_\odot}$ star, with initial metallicity $\mathrm{Z} = 0.02$ at the ZAMS stage.  The stellar structure includes a convective core that extends beyond the Schwarzschild boundary through a penetration zone, implemented using a simplified version of the convective boundary prescription described in \citet{Mao2024}. The implementation of this convective boundary treatment is evident in the temperature gradient ($\nabla$) profiles shown in Figure \ref{Fig:Kippenhahn} (right panel), which demonstrate the smooth transition of $\mathrm{\nabla_T}$ from the convective core through the penetration zone to the radiative envelope. The Kippenhahn diagram in Figure \ref{Fig:Kippenhahn} (left panel) marks the base state used in the simulations with the outer boundary $\mathrm{R_{max}}$ positioned at \unit{4100}{\Mm}, encompassing a hydrogen-burning convective core and intermediate radiative envelope. We employed a modified opacity model compared to the \texttt{MESA} opacity profile, with details provided in the following section.

\subsection{Modified opacity model}
\begin{figure}
   \centering
   \includegraphics[width=\hsize]{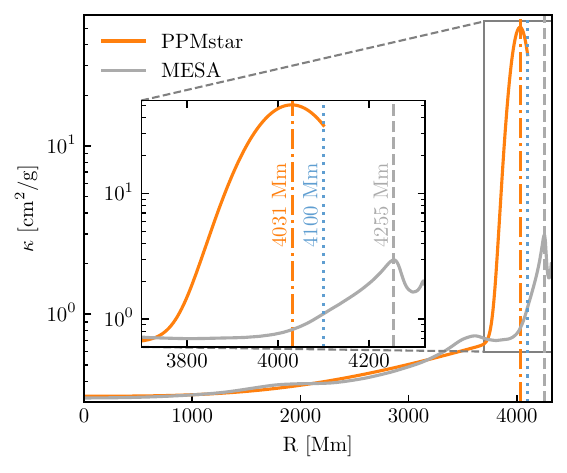}
      \caption{Comparison between the \texttt{MESA} opacity profile and the modified opacity model used in the \texttt{PPMstar} simulations. The grey dashed vertical line indicates the radial location of opacity bump maxima in the \texttt{MESA} profile, the orange dot-dashed line shows the radial location of the maxima in the modified opacity model, and the blue dotted line marks the outer boundary of the simulation domain.}
         \label{Fig:Opacity-model}
\end{figure}

\begin{figure}
   \centering
   \includegraphics[width=\hsize]{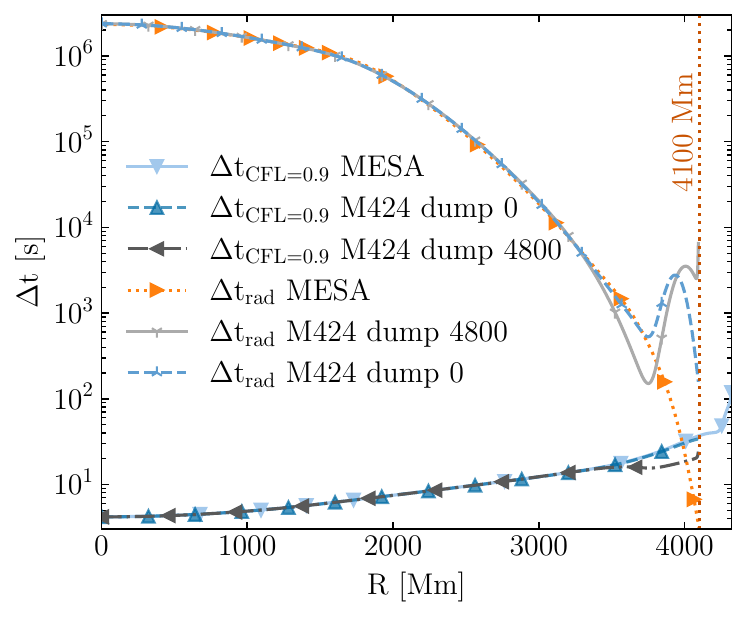}
      \caption{Comparison of CFL timestep \citep{Courant1928} (with Courant number 0.9) with radiation diffusion timestep \citep{Rider1999} for the \texttt{MESA} model, dump 0 and dump 4800 ($\approx$ 3779 h) of run M424 using spatial resolution $\Delta x$ = \unit{4.58}{\Mm}. The red dotted line marks the outer boundary at 4100 Mm.}
         \label{Fig:Timestep}
\end{figure}

Figure \ref{Fig:Opacity-model} compares the \texttt{MESA} opacity profile with the modified opacity model used in the simulations. We use this modified opacity profile boosted by a factor of 100 (100$\times$ radiative diffusivity). This boost matches the enhancement applied to core heating, preserving the global balance between convection and radiation in our simulations. The opacity remains unchanged throughout the short simulated time, and we use a fixed-in-time opacity profile. The opacity function $\kappa(r)$ for all runs is:
\begin{equation}
    \kappa(r) = p_0 + p_1\, \left( \frac{r}{1000\, \mathrm{Mm}}\right)^3 + p_2 \, \exp{\left(-\frac{(r-p_3)^2}{2\, p_4^2}\right)}
\end{equation}
with parameter values $p_0 = 0.33\;\mathrm{cm^2/g}$, $p_1 = 0.00658\;\mathrm{cm^2/g}$, $p_2 = 50.0 \;\mathrm{cm^2/g}$, $p_3$ = \unit{4030}{\Mm}, $p_4$ = \unit{80}{\Mm} and $r\;\mathrm{(Mm)}$ is the radial coordinate.  The \texttt{MESA} 1D model has a thin \unit{45}{\Mm} subsurface convection zone followed by a thin \unit{40}{\Mm} radiative zone. Three modifications were made to the opacity model as compared to \texttt{MESA} opacity profile. The Fe opacity bump, which produces the thin outer envelope Fe convection zone, was shifted inward from \unit{4255}{\Mm} to \unit{4030}{\Mm} to place it within the computational domain. The spread of this bump was increased to expand the radial extent of the convection zone, enabling better resolution of turbulent convection with more grid cells. The overall amplitude was increased by a factor of 17 to mitigate radiative diffusion timestep constraints, as discussed below.

We treat radiation in the diffusion limit. We adopt the radiative diffusion timestep $\Delta t_{\mathrm{rad}}$ from \citet{Rider1999}:
\begin{equation}
\Delta t_{\mathrm{rad}} = \frac{\Delta x^2}{4 \, \nu_{\mathrm{rad}}}
\end{equation}
where $\nu_{\mathrm{rad}} = \frac{2}{3R} \frac{k_{\mathrm{rad}}}{\rho}$ and $k_{\mathrm{rad}} = \frac{4\,a\,c\,T^3}{3\,\kappa\,\rho}$. Here $\Delta x$ is the grid spacing, $R$ is the gas constant, $\rho$ is the density, $a$ is the radiation constant, $c$ is the speed of light, $T$ is the temperature, and $\kappa$ is the opacity. Near the envelope, $\Delta t_{\mathrm{rad}}$ for \texttt{MESA} becomes comparable to and even smaller than the CFL timestep limit \citep{Courant1928}, as shown in Figure \ref{Fig:Timestep}. We require $\Delta t_{\mathrm{rad}} > \Delta t_{\mathrm{CFL}}$ to take sufficiently large timesteps within the CFL limit for long duration simulations needed to perform time series analysis. The \texttt{MESA} densities drop rapidly in the outermost layers, and the timestep in those layers becomes prohibitively limited by the radiation timescale because we do explicit diffusion. Reducing the timestep to extend further outward is ineffective because the opacity drops steeply in the outermost layers. This requires us to choose $\rm R_{max}$ such that the radiation timestep constraint is satisfied with our CFL condition. Therefore we choose $\rm R_{max}$ in our simulations such that we can resolve the diffusion in the outermost layers with our modified opacity model. As shown in Figure \ref{Fig:Timestep}, the radiative diffusion timestep limit in M424 remains above the CFL timestep limit (using Courant number 0.9). This characteristic applies to all runs as the same opacity model is used throughout.

\begin{figure}
   \centering
   \includegraphics[width=\hsize]{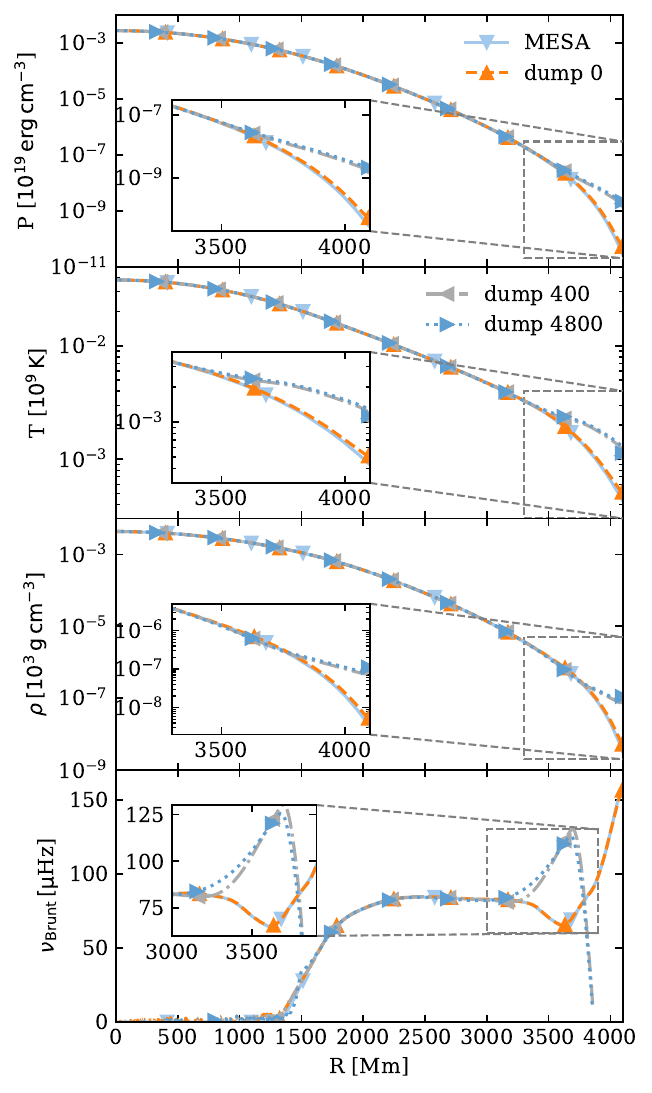}
      \caption{Comparison of pressure ($\mathrm{P}$), temperature ($\mathrm{T}$), density ($\mathrm{\rho}$), and linear Brunt--Väisälä frequency ($\mathrm{\nu_{Brunt}}$) stratifications between the \texttt{MESA} state, dump 0, dump 400 ($\approx$ \unit{78}{\hour} and dump 4800 ($\approx$ \unit{3779}{h}) of run M424. All panels share the same legend as shown in the top two panels.}
         \label{Fig:Stratification-comparison}     
\end{figure}

Figure \ref{Fig:Stratification-comparison} shows eight orders of magnitude difference in pressure, four orders of magnitude difference in temperature, and six orders of magnitude difference in density between the \texttt{MESA} profile and dump 0. Although the outer convection zone is convectively unstable, radiation remains an efficient heat transfer mechanism there. Therefore, we observe that $\nabla$ follows $\nabla_{\mathrm{rad}}$ around the opacity bump (Figure~\ref{Fig:Kippenhahn}). The Brunt--Väisälä frequency remains positive in the outer convection zone because radiation effectively transports heat, maintaining stability against buoyancy perturbations despite the convective instability.

\begin{table*}
\centering
\caption{Summary of simulation parameters}
\label{tab:simulation_params}
\begin{tabular}{lccccccr}
\hline\hline
Run ID & Configuration & Grid & Core Heating & $R_{\text{min}}$ & $R_{\text{max}}$ & Cadence & Total simulation\\
 &  & resolution & enhancement factor & (Mm) & (Mm) & (min) & time (h)\\
\hline
M424 & Fullstar run & $1792^3$ & 100 & 0    & 4100 & 47 & 3779 \\
M438 & No core heating & $1728^3$ & 0   & 0    & 4100 & 46 & 3947 \\
M484 & Outer convection shell & $1792^3$ & 100 & 3850 & 4100 & 44 & 2603\\
M487 & No outer convection & $1792^3$ & 100 & 0    & 3600 & 41 & 3702\\ 
 & (large radiative envelope) &  &  &     &  &  & \\
M488 & No outer convection & $1792^3$ & 100 & 0    & 2670 & 45 & 2824\\ 
 & (small radiative envelope) &  &  &     &  &  & \\
\hline
\end{tabular}
\end{table*}
The differences between dump 0 and dump 400 in Figure \ref{Fig:Stratification-comparison} demonstrate the fast initial thermal readjustment of the stellar structure in response to the modified opacity profile. The \texttt{PPMstar} simulation begins with the ZAMS \texttt{MESA} stratification but immediately starts evolving toward a new thermal equilibrium consistent with the enhanced and repositioned Fe opacity bump. This rapid thermal adjustment, occurring over approximately 310 h (much shorter than the stellar thermal timescale), reflects the star's response to the artificially modified opacity structure and results in the equilibrated stratification profiles shown at dump 400. Therefore, while our initial base state represents 95\% of the stellar radius, the outermost 8\% layers of the simulated star exhibit approximately ten times higher density than the MESA stellar model. This discrepancy results from the implementation of a convection zone in the outermost layer that is both thicker and positioned deeper than in the actual star. A stellar model with the convection zone present in our equilibrated hydrodynamic setup would likely have a larger stellar radius. The \texttt{MESA} ZAMS model has no mean molecular weight ($\mu$) gradient except near the outer boundary due to partial ionization of chemical species as temperature decreases toward the surface. This effect is ignored in the initial setup of all runs, resulting in a flat $\mathrm{\mu = 0.617317}$ mean molecular weight profile.
\subsection{Luminosity post-processing and spectral analysis}
\texttt{PPMstar} simulations save two different types of compressed data at equal intervals of time (dumps) which are substantially larger than the simulation time steps. These include \texttt{rprof}, which contains spherically averaged radial profiles at full grid resolution, and \texttt{briquette}, which stores 4x compressed 3D data as described in \citet{Stephens2021}.

Temporal mock spectra are generated from the line-of-sight hemispherically averaged light curves at a given radius as described in \citet{Thompson2024}. These mock light curves in eight different lines-of-sight are calculated in-line in the code at full grid resolution and output in the \texttt{rprof} files. These light curves are detrended with a $3^{\rm rd}$-order polynomial to remove the global trend, unity subtracted relative luminosity is computed, individual power spectra are calculated, and the results from all eight different lines of sight are averaged.

The power spectra are fitted with a Lorentzian function \citep{Blomme2011,Bowman2019a,Bowman2019b}: 

\begin{equation}  
\alpha(\nu) = \frac{\alpha_0}{ 1 + \left(\frac{\nu}{\nu_{\text{char}}}\right)^\gamma} + \mathrm{C_w},
\end{equation} 
where $\alpha(\nu)$ represents the Lorentzian curve in $\mathrm{\mu Hz}$, $\alpha_0$ is the amplitude at zero frequency (in $\mathrm{\mu Hz}$), $\nu$ and $\nu_{\text{char}}$ are the frequency and characteristic frequency respectively (both in $\mathrm{\mu Hz}$), $\gamma$ is the dimensionless logarithmic amplitude gradient, and $\mathrm{C_w}$ denotes the frequency independent noise (in $\mathrm{\mu Hz}$).

We use the 3D \texttt{briquette} data at a spherical shell of fixed radius $R$ to construct the $\ell-\nu$ diagrams. Here $\ell$ is the spherical harmonic order and $\nu$ is the cyclic frequency. The $\ell-\nu$ diagrams are created for the calculated variable `unity-subtracted relative luminosity' $\mathcal{L}$ \citep[same as in][]{Thompson2024}, using the \texttt{briquette} temperature data ($\mathrm{T}$) as:
\begin{align}
    \mathcal{L} = \mathrm{\frac{ T^4}{ \langle T^4\rangle}} - 1,
\end{align}
where $\mathrm{\langle T^4\rangle}$ is the spherical average of 3D \texttt{briquette} temperature variable, at the radius of interest, which is used as a base to remove the global luminosity trend.

Table \ref{tab:simulation_params} summarizes all runs with their respective simulation parameters.  These configurations enable systematic investigation of the individual and combined contributions of different stellar regions to the observed SLF variability. For the runs presented in this study, a modified opacity model is used as compared to the \texttt{MESA} opacity profile. Compared to the \texttt{MESA} model, core heating (mimicking core H burning) and thermal diffusivity are boosted by a factor of 100 to achieve numerically tractable fluid velocities and to scale down thermal diffusion timescales, respectively. 

Several numerical constraints determine the choice of enhancement factor for core heating and thermal diffusivity. Minimizing the Mach number in the thin outer envelope convection zone requires making thermal diffusivity large enough to reduce flow velocities, yet small enough to preserve convective instability necessary for convection to occur. We boost thermal diffusivity and luminosity by the same factor to maintain global thermal equilibrium and the stratification. Luminosity and radiation diffusion are boosted by the same factor to accelerate the stellar evolution simulation while preserving the underlying physics. This balanced enhancement speeds up both the thermal timescale ($\propto$ boost factor$^{-1}$) and convective timescale ($\propto$ boost factor$^{-1/3}$), making the ratio of thermal to dynamic timescale smaller with larger boost factors and thus computationally feasible to reach thermal and dynamic equilibrium \citep{Mao2024}. This optimization imposes the most restrictive constraint on the thermal diffusion timestep limit, necessitating the coordinated adjustment of both the boost factor and grid resolution. These competing numerical requirements lead to the $100\times$ boost factor for core heating and thermal diffusivity enhancement. The combination of $1792^3$ resolution with the 100× boost factor represents an optimal configuration that satisfies these constraints. Previous convergence studies by \citet{Herwig2023}, \citet{Thompson2024}, and \citet{Mao2024} demonstrate that this resolution is high enough to achieve convergence for convective dynamics, entrainment rates, and envelope vorticity in 3D stellar convection simulations.

For each simulation, several core-convective turnover times are allowed to elapse before beginning analysis, ensuring fully developed convection in both the core (218 h turnover) and the envelope (3 h turnover).

\section{Results}
\label{sec:Results}
\subsection{Fullstar run}
\begin{figure*}
\begin{interactive}{animation}{animations/M424-merged_movie.mp4}
   \centering
   \includegraphics[width=\hsize]{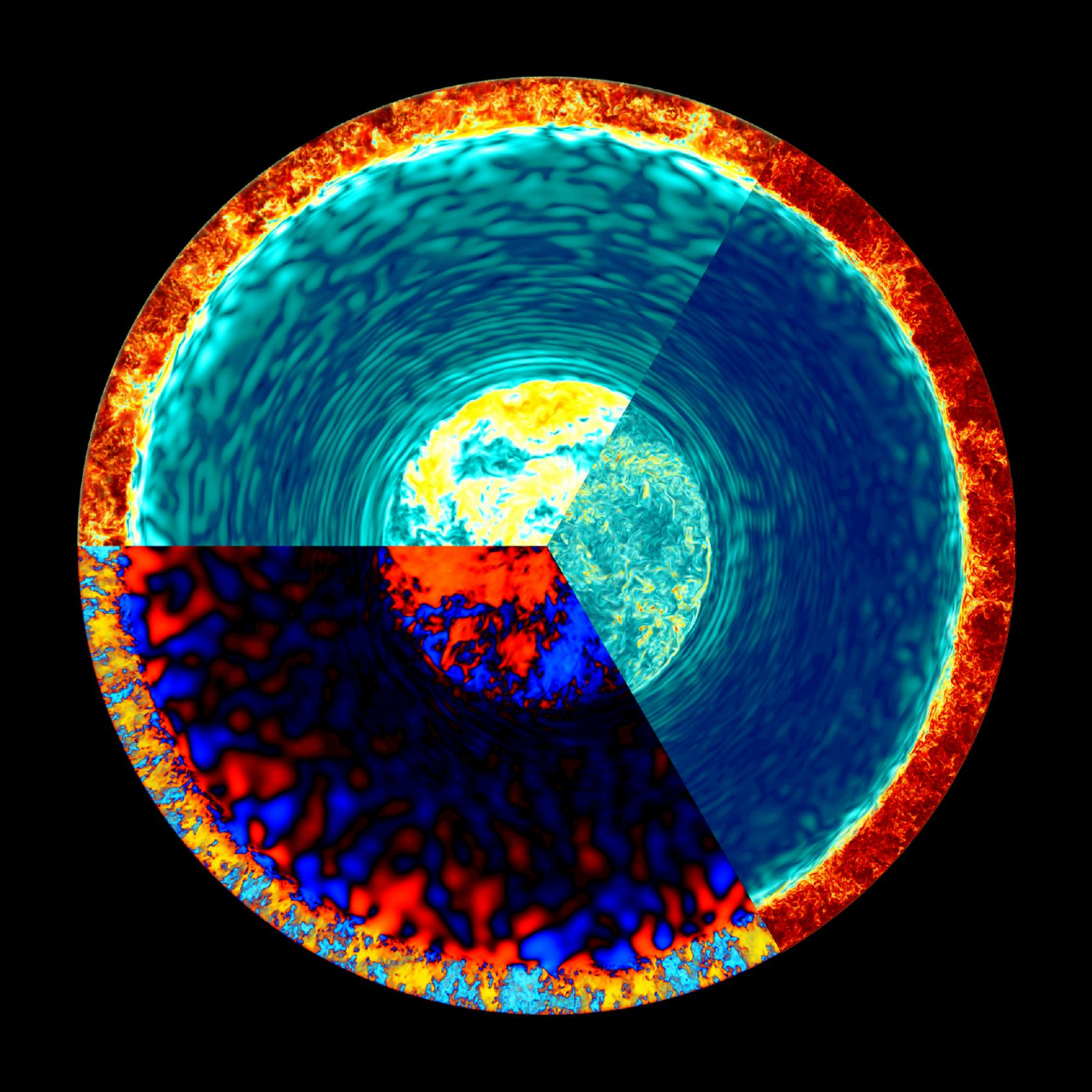}
   \put(-410,445){\colorbox{black}{\textcolor{white}{\textbf{(I)}}}}
   \put(-30,250){\colorbox{black}{\textcolor{white}{\textbf{(II)}}}}
   \put(-375,40){\colorbox{black}{\textcolor{white}{\textbf{(III)}}}}
\end{interactive}
   \caption{Volume-rendered visualization of three fluid variables for the run M424 at dump 2000 ($\approx 1990$ h), generated using a thin equatorial slice: (I) Horizontal velocity magnitude, represented by a color gradient from highest to lowest: dark brown, red, yellow, white, light blue, and dark blue; (II) Vorticity magnitude, depicted with a color scheme from highest to lowest: red, yellow, light blue, and dark blue; (III) Radial velocity, where inward-directed (negative) velocities are shown in light to dark blue (decreasing magnitude), and outward-directed (positive) velocities are shown in yellow, orange, and red (decreasing magnitude). An animation is available in the HTML version showing the temporal evolution of these quantities over 30 consecutive dumps.}
   \label{Fig:M424-combined}
\end{figure*}
Figure~\ref{Fig:M424-combined}\footnote{Volume-rendered visualizations of various fluid variables for all simulations presented in this paper are available at \url{https://www.ppmstar.org/}.} shows the turbulent core and thin outer envelope convection, separated by an intermediate radiative zone where wave-like features corresponding to internal gravity waves (IGWs) are evident. The turbulent convective core extends from the center to $\approx$ \unit{1500}{\Mm}, exhibiting large-scale dipole circulation patterns visible in the radial velocity and various small-scale turbulent features apparent in the vorticity magnitude. Thin ring-like IGW features are visible in the vorticity magnitude and horizontal velocity, originating near the core convective boundary in the radiative zone from $\approx$ \unit{1500}{\Mm} to \unit{3733}{\Mm}. A strong thin outer envelope convection zone appears as a thick ring around the outer boundary from $\approx$ \unit{3733}{\Mm} to \unit{4100}{\Mm}, displaying high relative magnitudes in all three variables. A non-linear color mapping was employed to highlight these multi-scale features. The convective core morphology is discussed in detail in our previous papers \citep{Herwig2023,Thompson2024,Mao2024}. 
\begin{figure}
   \centering
   \includegraphics[width=\hsize]{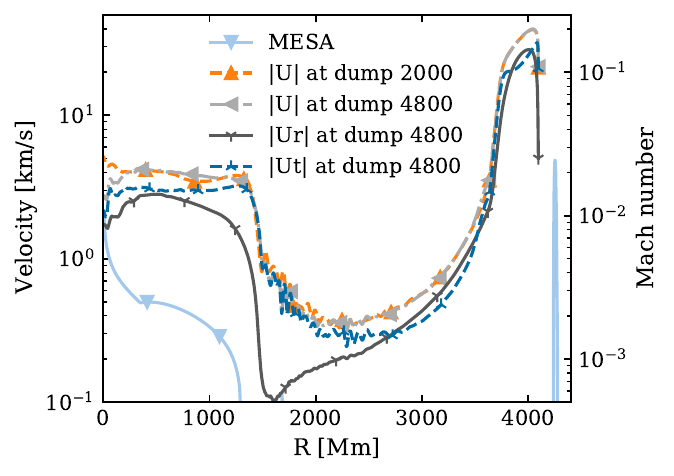}
      \caption{Comparison of \texttt{MESA} convective velocity profile with velocity magnitude $|U|$ for run M424 at dumps 2000 and 4800, and radial $|U_r|$ and tangential $|U_t|$ velocity magnitudes for M424 at dump 4800. The secondary y-axis shows the Mach number for the M424 profiles.}
         \label{Fig:FeCZ-convective-velocities-M424}
\end{figure}

\begin{figure}
   \centering
   \includegraphics[width=\hsize]{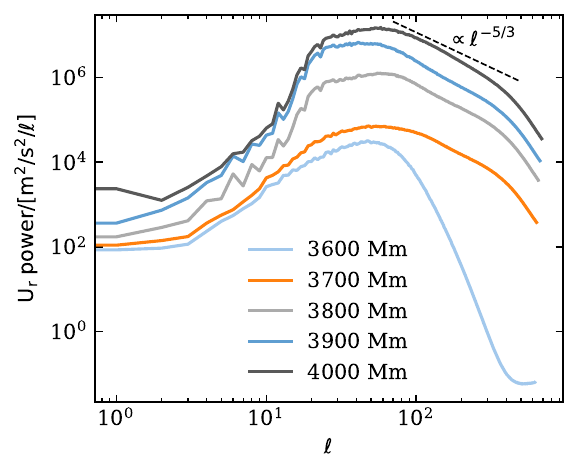}
      \caption{Radial velocity spatial spectra of the run M424 at different radial coordinates near the surface averaged over last 100 dumps.}
         \label{Fig:Spatial_spectra-M424}
\end{figure}

The modified opacity model (Figure~\ref{Fig:Opacity-model}) implemented in our simulations produces strong turbulent thin outer envelope convection with turnover timescale of $\approx 3$ h. Figure~\ref{Fig:FeCZ-convective-velocities-M424} shows that we see a factor of $\approx$ 10 higher core convective velocities in M424 compared to \texttt{MESA} MLT velocities. Previous work by \citet{Herwig2023} shows that core convective velocities scale with luminosity boost factor as boost factor$^{1/3}$. This means at nominal luminosity, the core convective velocities are 4.64 times smaller but still 2-3 times larger than \texttt{MESA} MLT velocities. This is consistent with \citet{Jones2017} who found $\approx 2$ factor higher velocities as compared to \texttt{MESA} MLT velocity in their 3D \texttt{PPMstar} simulations of turbulent oxygen-burning shell convection at nominal luminosity. Figure~\ref{Fig:FeCZ-convective-velocities-M424} also shows that convective velocities in the thin outer envelope zone are five to eight times larger than the thin outer envelope convection in the \texttt{MESA}. Figure~\ref{Fig:Spatial_spectra-M424} confirms that the radial velocity $\mathrm{U_r}$ spatial power spectrum inside the thin outer envelope convection zone follows the Kolmogorov's power law of $\ell^{-5/3}$, indicating that spatial scales at which most of the power is added to the turbulent convection are well separated and larger than the spatial scales at which energy is dissipated into heat. This figure also indicates that the dominant power is concentrated around the spatial scale of $\ell \approx 40$. This spatial scale corresponds to the radial extent of the thin outer envelope convection zone.

The thermal diffusion timescale of the thin outer envelope convection zone for run M424 is 142 h. This is computed by integrating the $\rm \Delta t_{rad}$ curve over the thin outer envelope convection zone. We assumed the radial extent of the thin outer envelope convection zone from \unit{3733}{\Mm} (peak of Skew $\times$ Kurtosis, see Appendix~\ref{app:convective-boundary}) to the outer boundary \unit{4100}{\Mm}. Thermal equilibrium gradually moves inward, initially fast, then slowing down as thermal timescales increase with depth in the star. The thermal timescale of the inner radiative region from \unit{1500}{\Mm} to \unit{2500}{\Mm} is $\approx 10^6$ h. We ensure that the outer layers of interest are not changing significantly over the duration needed to create spectra (Figure~\ref{Fig:Stratification-comparison}). We determine the time interval required to make spectra and wait long enough so that the stratification remains stable over this analysis period. Therefore, we choose to analyze run M424 starting from dump 2000 ($1990$ h).

\begin{figure}
   \centering
   \includegraphics[width=\hsize]{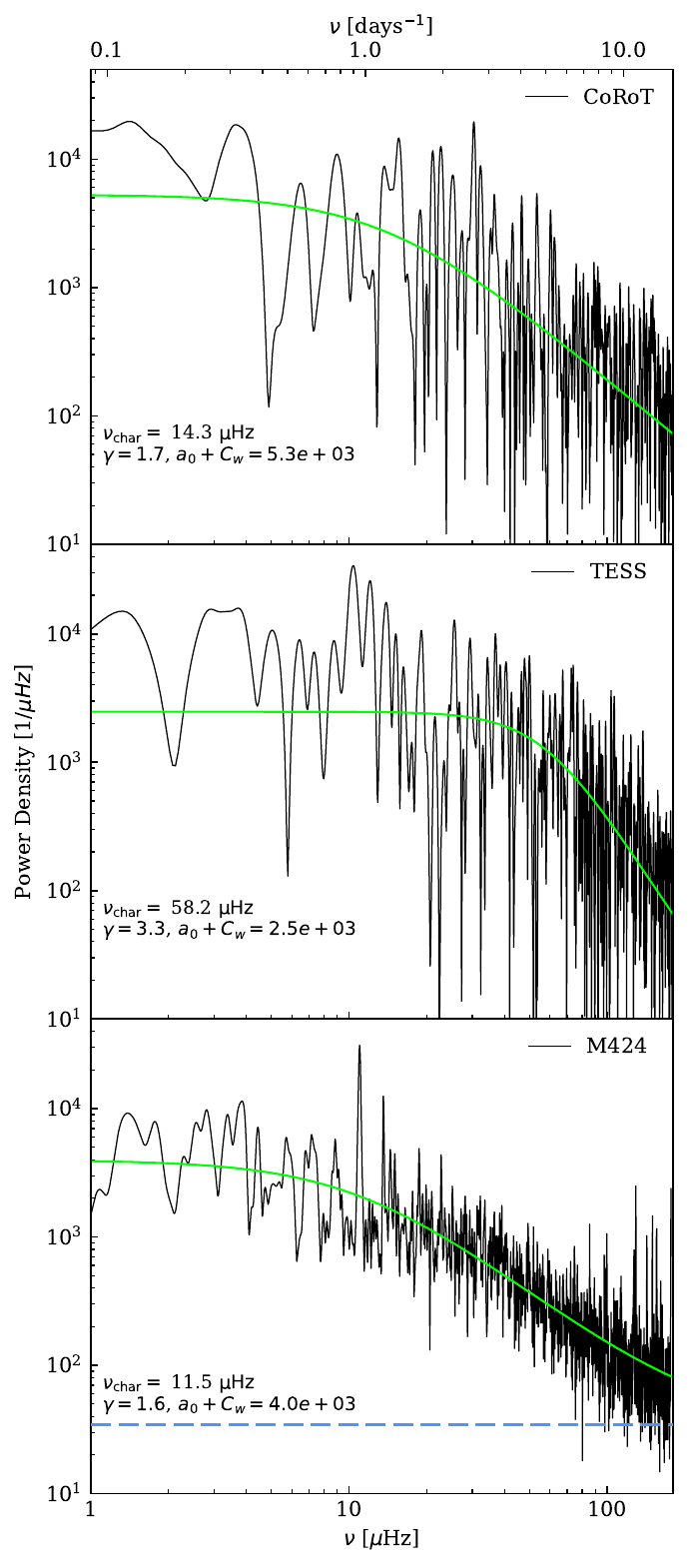}
   \caption{Comparison of luminosity power spectra between CoRoT and TESS observations of the O-star HD46150 and M424 simulation at \unit{4000}{\Mm}. The lime curve represents the best-fit semi-Lorentzian function, and the best-fit parameters are shown in the respective plots.}
   \label{Fig:observed-spectra}
\end{figure}

Figure~\ref{Fig:observed-spectra} compares the luminosity power spectrum of run M424 near the surface (\unit{4000}{\Mm}) with observations of the O-star \texttt{HD46150}\footnote{\texttt{HD 46150} is a young main-sequence O dwarf \citep{Bowman2019a} that is close to our ZAMS simulations.} using CoRoT (Convection, Rotation and planetary Transits) and TESS (Transiting Exoplanet Survey Satellite) light curves \citep{Bowman2019a,Bowman2020}.  We applied the same detrending and Fourier transform algorithms to the observed light curves as used for our simulated mock light curves. The TESS light curves have 21.77 days duration with 2-minute cadence, corresponding to a frequency range of 0.5 to $4.1\times10^3\;\mu\mathrm{Hz}$. The CoRoT light curves span the same duration but with variable cadence ranging from 0.5 to 20 minutes due to instrumental periodicities caused by the satellite's low-Earth orbit that create power artifacts in the spectrum. The mock M424 light curve spans 2800 dumps ($\approx$ \unit{2193}{\hour}) with 47-minute cadence, yielding a frequency range of 0.12 to $177\;\mu\mathrm{Hz}$. We therefore restrict the spectral plots in Figure~\ref{Fig:observed-spectra} to the frequency range relevant to M424 for comparison. The earlier study of \citet{Thompson2024} showed qualitative similarities with observations but exhibited quantitative differences, such as $\nu_{\mathrm{char}} \approx 6~\mu$Hz, which is an order of magnitude smaller than observations. After incorporating a larger radiative envelope and a (modified) thin outer envelope convection zone in this study, the spectrum of run M424 demonstrates both qualitative and quantitative similarities with observations. All three panels display approximately a two-order-of-magnitude difference between power at 180 $\mu$Hz and 1 $\mu$Hz, with $\nu_{\mathrm{char}}$ values of the same order of magnitude as observed spectra. The difference in Lorentzian fit parameters between TESS and CoRoT observations exceeds the difference between the simulation and TESS observation, which is considered more reliable. The dynamic range of spectral features spans several orders of magnitude in both TESS and M424 spectra, although spectral troughs are deeper in the TESS spectrum.

Given that run M424 incorporates a convective core, an intermediate radiative zone, and a convective envelope, the question arises regarding the relative contribution of convection and IGWs generated at both the core and envelope convective boundaries to the observed spectra. Are the features, such as peaks and troughs, observed in the simulations related to eigenmodes?
\begin{figure}
\begin{interactive}{animation}{animations/M438-merged_movie.mp4}
   \centering
   \includegraphics[width=\hsize]{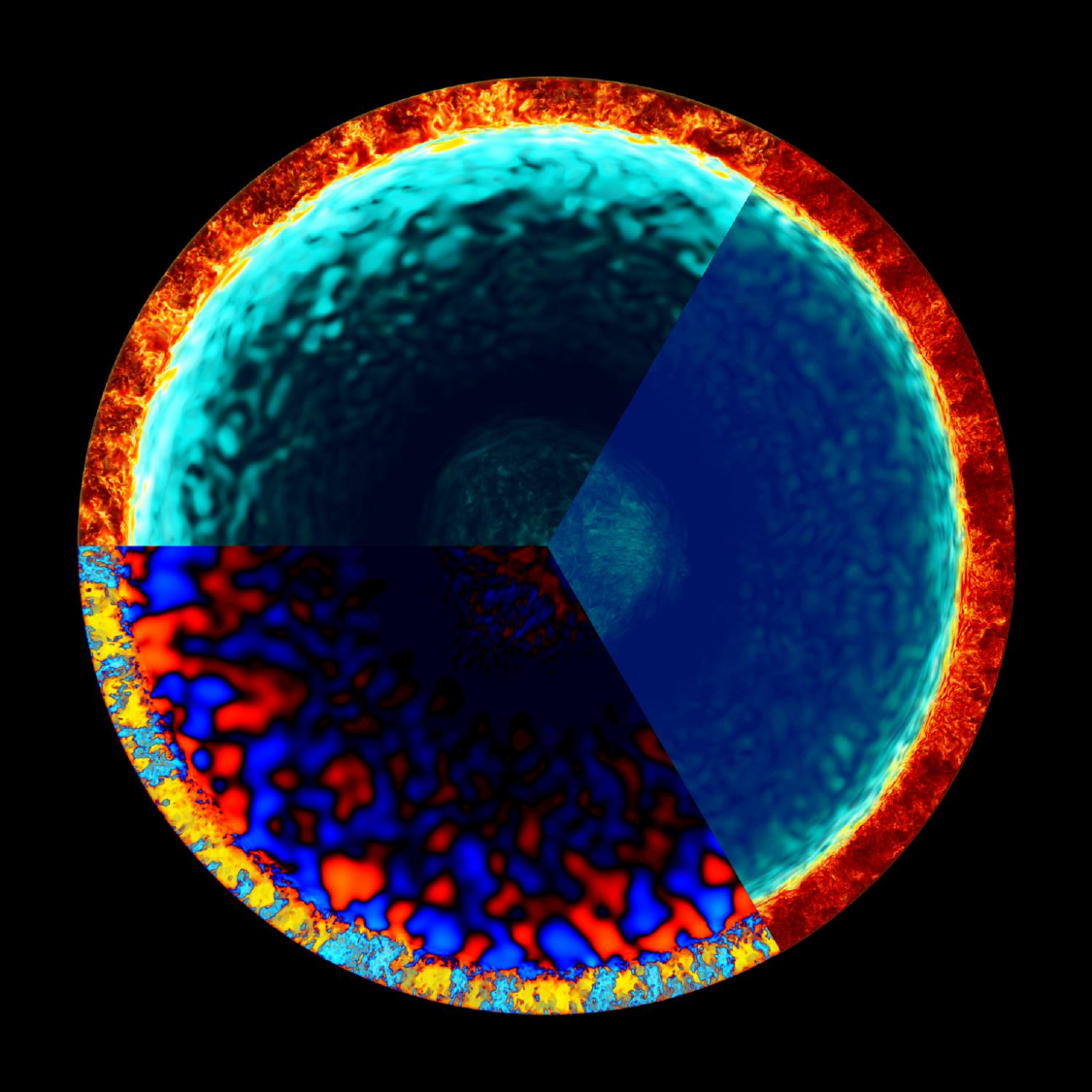}
\end{interactive}
   \put(-410,220){\colorbox{black}{\textcolor{white}{\textbf{}}}}
   \caption{Volume rendered images similar to Figure~\ref{Fig:M424-combined} but for the run M438. An animation is available in the HTML version showing the temporal evolution over 30 consecutive dumps.}
   \label{fig:M438-combined}
\end{figure}
\begin{figure}
\begin{interactive}{animation}{animations/M484-merged_movie.mp4}
   \centering
   \includegraphics[width=\hsize]{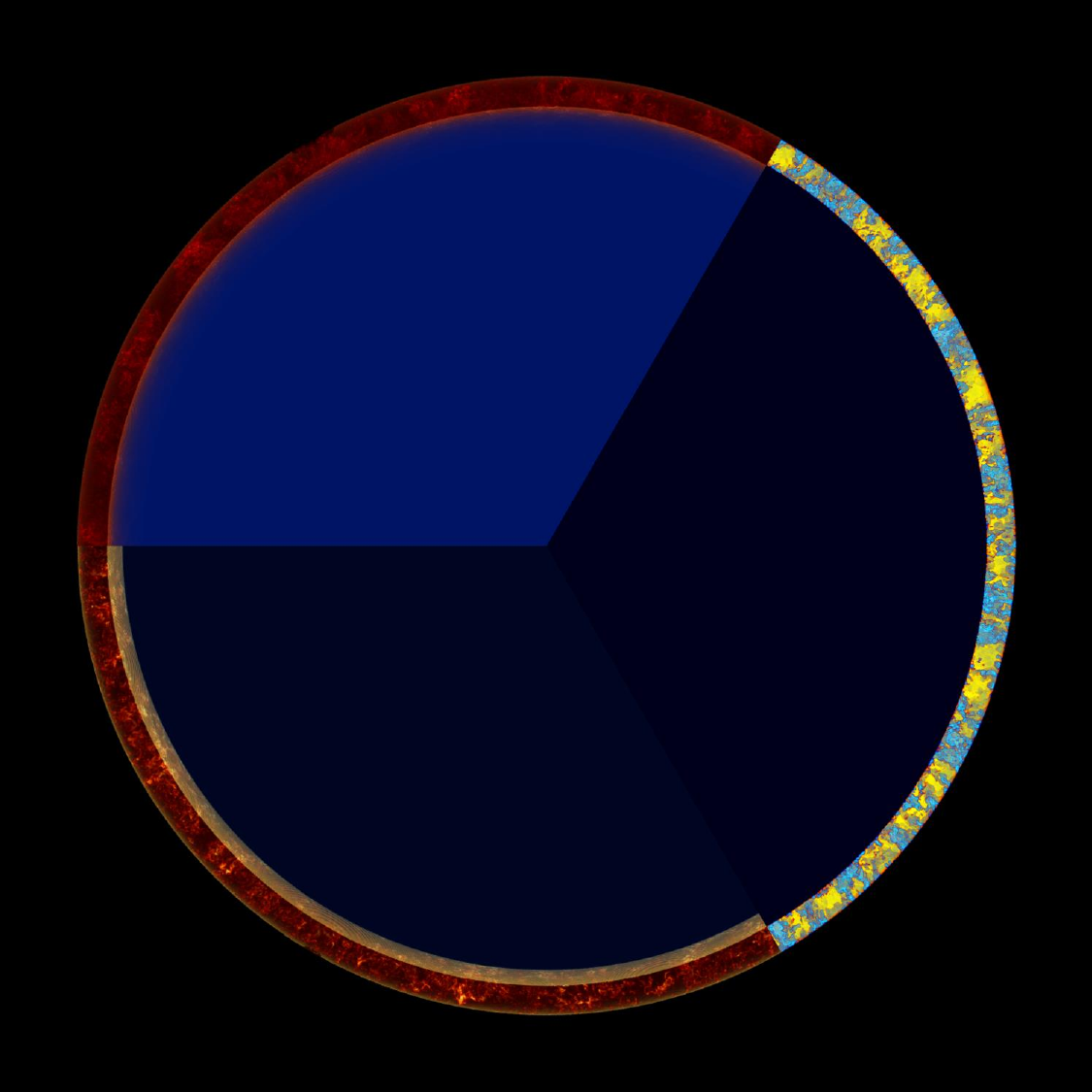}
\end{interactive}
   \put(-410,220){\colorbox{black}{\textcolor{white}{\textbf{}}}}
   \caption{Volume rendered images similar to Figure~\ref{Fig:M424-combined} but for the run M484. An animation is available in the HTML version showing the temporal evolution over 30 consecutive dumps.}
   \label{fig:M484-combined}
\end{figure}
\begin{figure}
   \centering
   \includegraphics[width=\hsize]{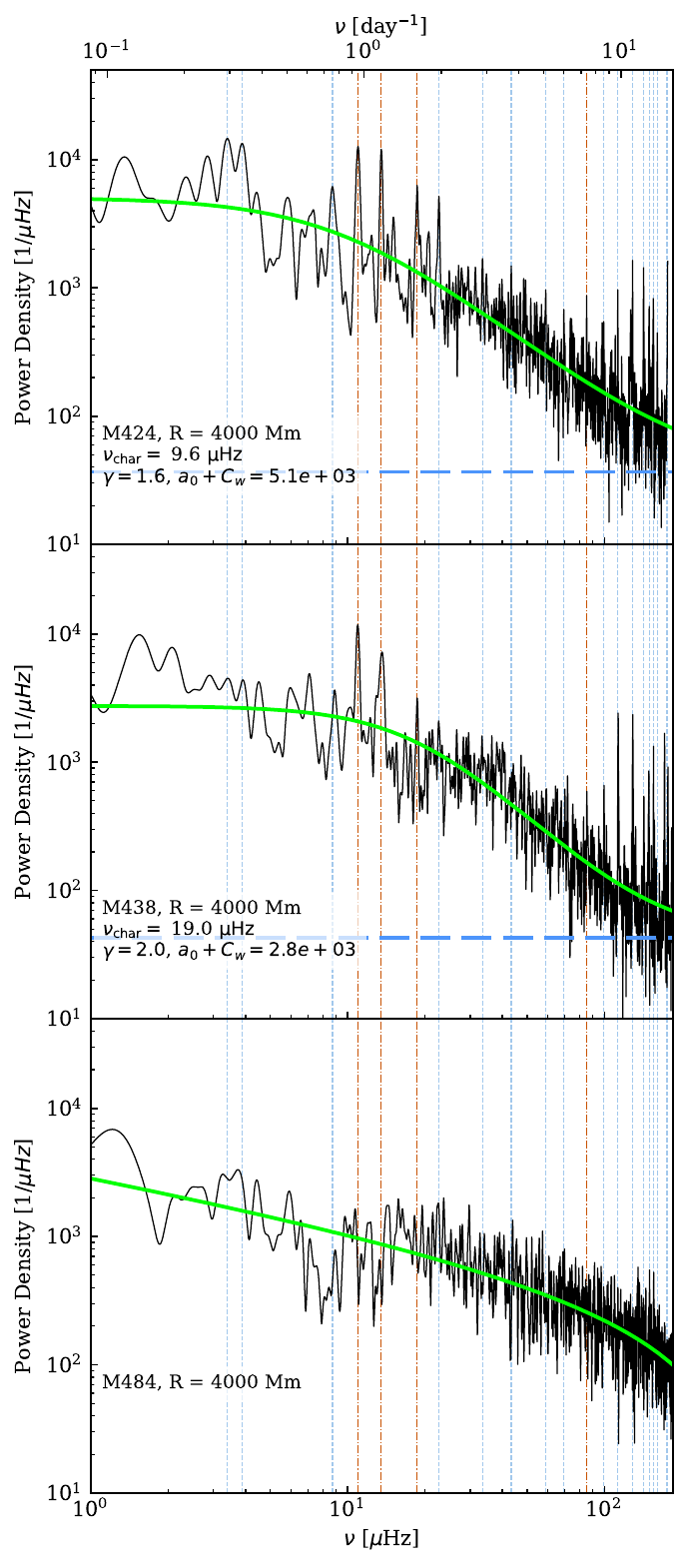}
   \caption{Comparison of luminosity power spectra at \unit{4000}{\Mm} across different simulation runs. Each spectrum was generated using the final 1600 dumps ($\mathrm{\approx 1200\; h}$) of simulation data. We print the name of the run, the radius at which the spectrum was made and the best-fit Lorentzian parameters for respective panels with exception to last panel for M484 where we could not find a good Lorentzian fit. We plot several vertical dotted lines in all the subplots marking the sharp features present in M438 and M424.}
   \label{Fig:power-spectra-4000Mm}
\end{figure}
\begin{figure}
   \centering
   \includegraphics[width=\hsize]{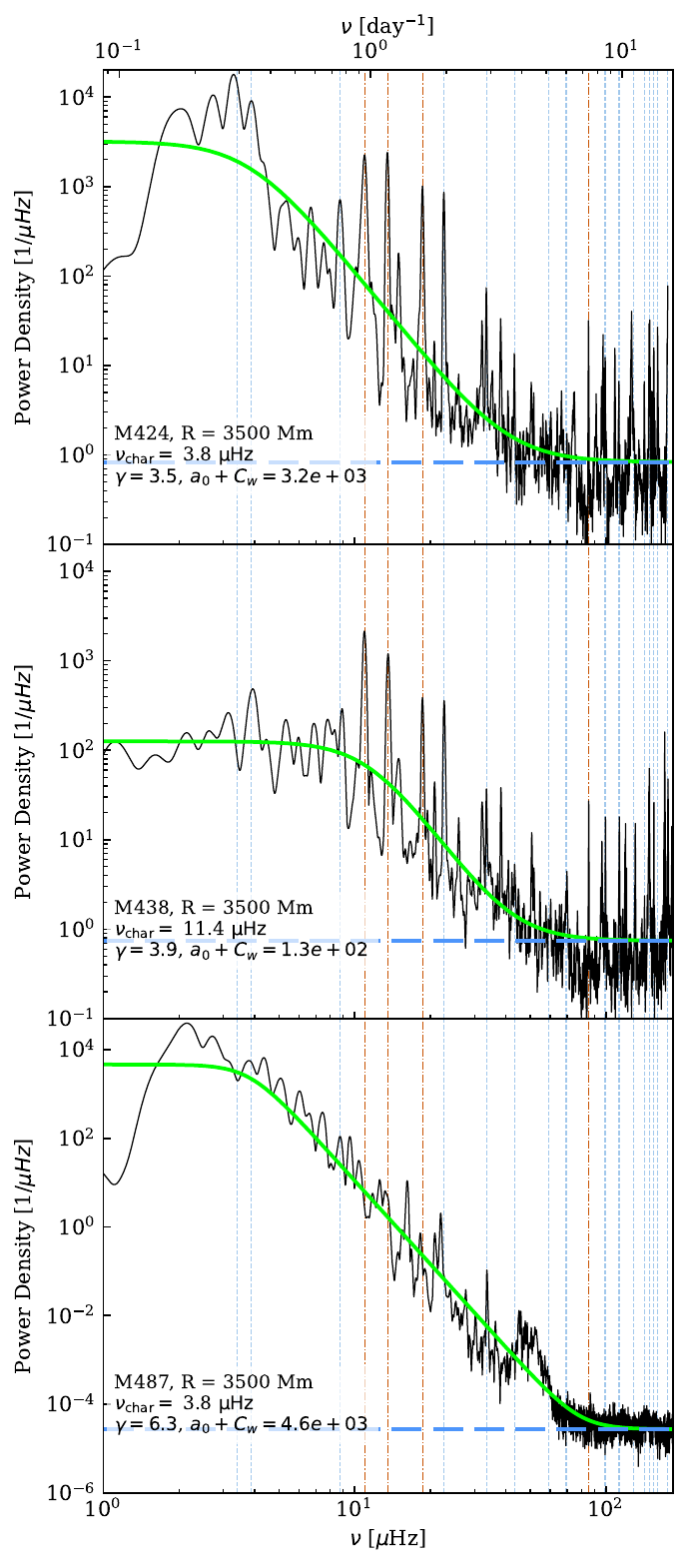}
   \caption{Same as Figure~\ref{Fig:power-spectra-4000Mm} but the luminosity power spectra made at \unit{3500}{\Mm} across different simulation runs. Same vertical dotted lines are plotted as in Figure~\ref{Fig:power-spectra-4000Mm}.}
   \label{Fig:power-spectra-3500Mm}
\end{figure}
\subsection{No core heating} Run M438 is identical to M424 but omits constant
volume heating in the center that in M424 represents the core H burning. This
means that as before the entire core is included but because no heating is applied the
initial constant-entropy profile in the core does not change and stays isentropic throughout the
simulation. This modification is evident in Figure~\ref{fig:M438-combined}, which shows
the absence of turbulent core convection compared to Figure~\ref{Fig:M424-combined}. This
experiment was designed to investigate how eliminating core convection, and consequently
the IGWs excited at the core-convective boundary, affects the frequency spectrum,
particularly the low-frequency excess.

In Figure~\ref{Fig:power-spectra-4000Mm}, we plot the luminosity power spectrum inside the thin outer envelope convection zone at \unit{4000}{\Mm}. For direct comparison between runs, we use the final 1600 dumps from each simulation, as not all runs have sufficient duration for longer time series (Table~\ref{tab:simulation_params}). With a cadence of $\approx 45$ minutes, this yields frequency range from 0.2 to 185 $\mu$Hz in the luminosity power spectrum. The luminosity power spectra reveal approximately equivalent power in the convective envelope when comparing M424 (with core convection) and M438 (without core convection) at \unit{4000}{\Mm}. 
Following \citet{Pedersen2025}, we quantitatively compare the two simulated spectra using the integrated total power as a model-independent parameter and find that it is 10\% less in M438 than in M424. This means that the convective core and the IGWs originating at the convective core boundary contribute only a small (10\%) effect to the low-frequency excess.

In Figure~\ref{Fig:power-spectra-4000Mm}, several marked spectral features appear at identical frequencies in both M424 and M438, which cannot be attributed to convection due to its stochastic nature. The same features appear in the spectral analysis at \unit{3500}{\Mm} for both runs in Figure~\ref{Fig:power-spectra-3500Mm}. The location at \unit{3500}{\Mm} is within the radiative zone below the thin outer envelope convection boundary and beyond the convective overshoot region (Appendix~\ref{app:convective-boundary}). Therefore at \unit{3500}{\Mm}, the spectrum and its features must be due to IGWs. Since the same features appear at \unit{4000}{\Mm}, they must originate from evanescent IGWs in the thin outer envelope convection zone and the spectrum in these simulations is dominated by IGWs.

\begin{figure*}
% First row (your existing figures)
\includegraphics[width=0.49\textwidth]{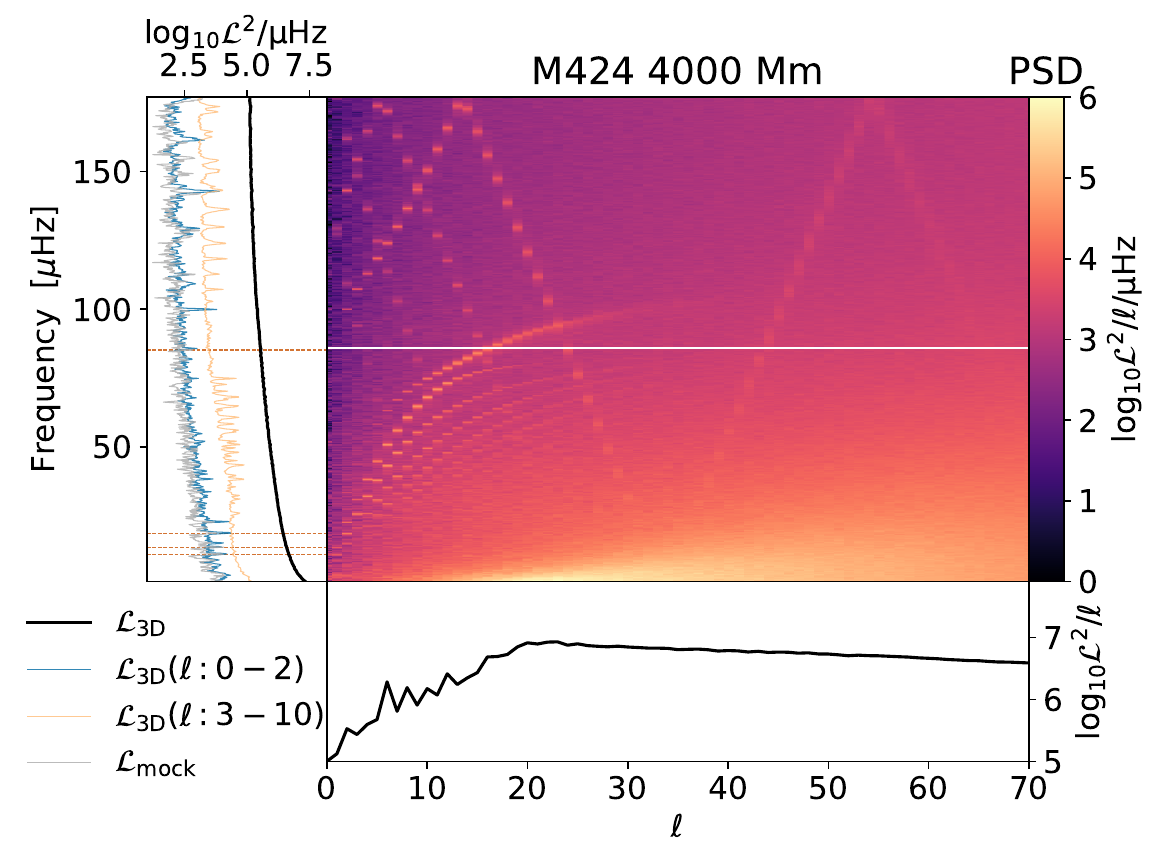}
\includegraphics[width=0.49\textwidth]{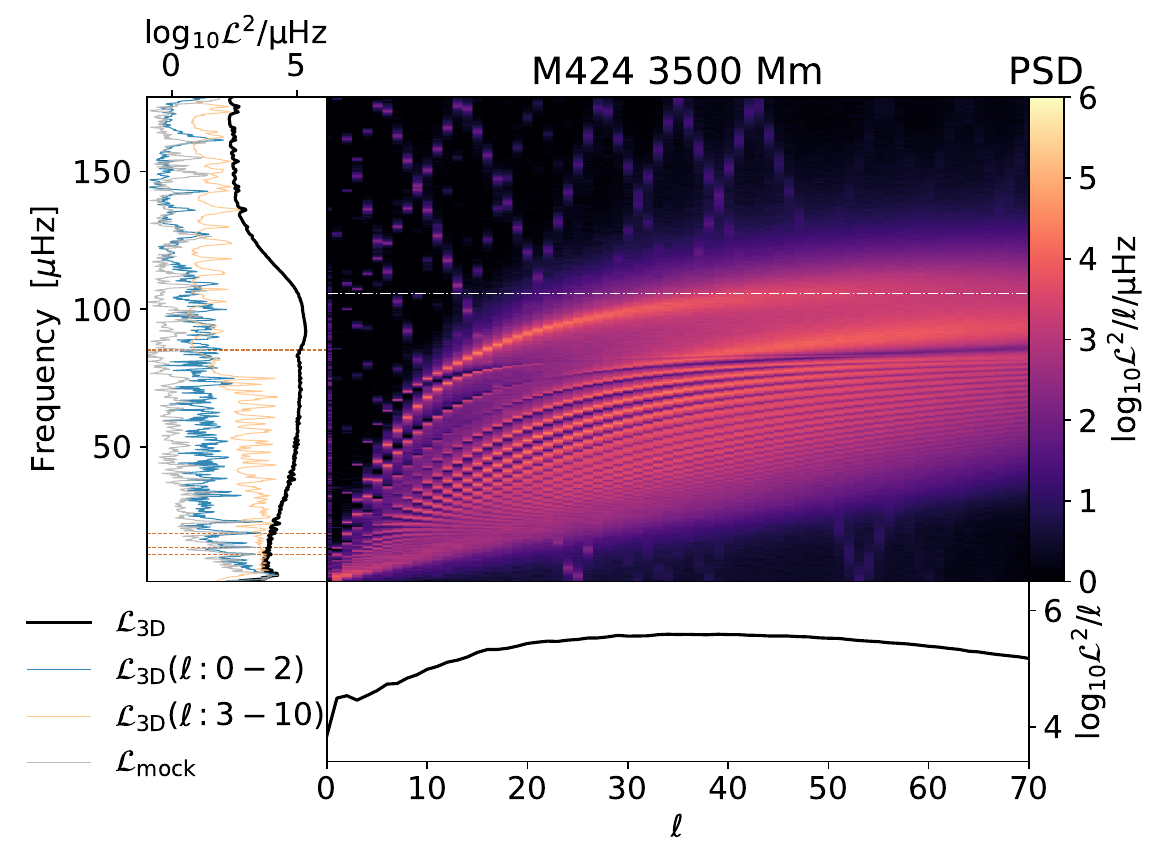}

% Second row
\includegraphics[width=0.49\textwidth]{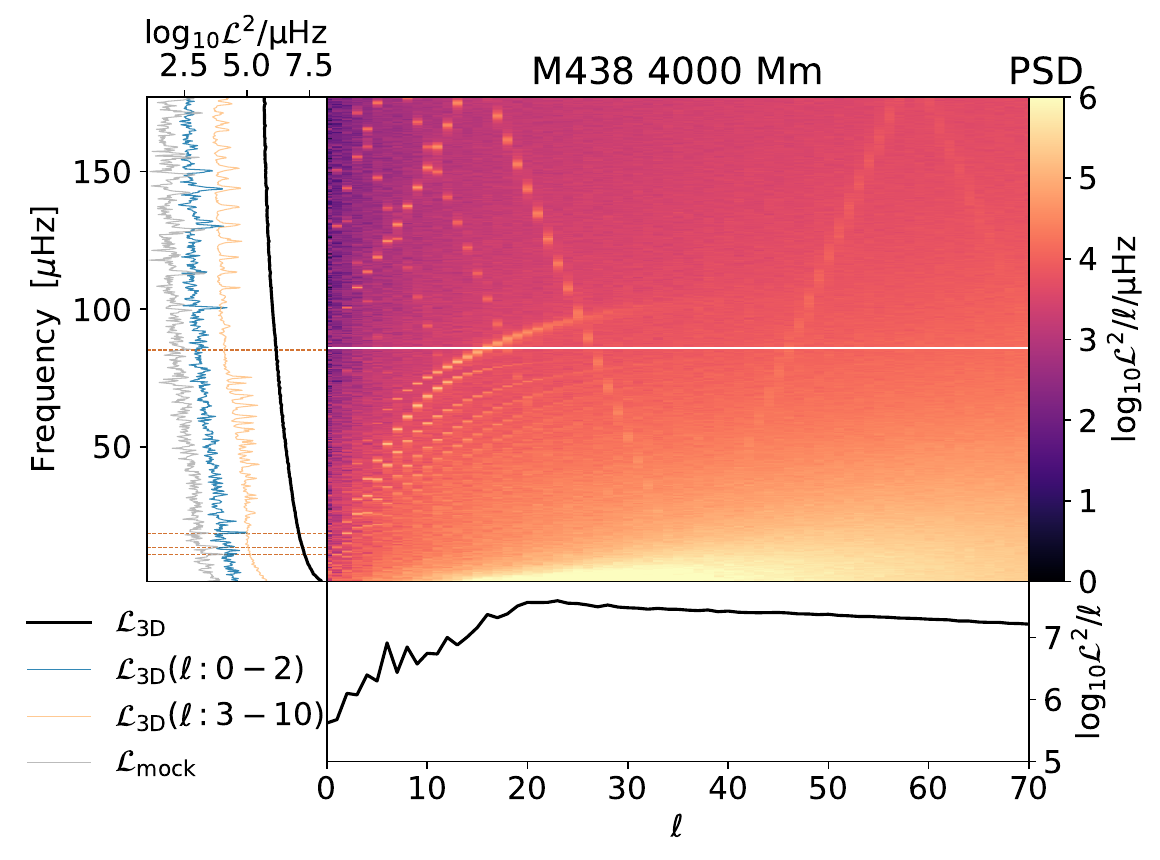}
\includegraphics[width=0.49\textwidth]{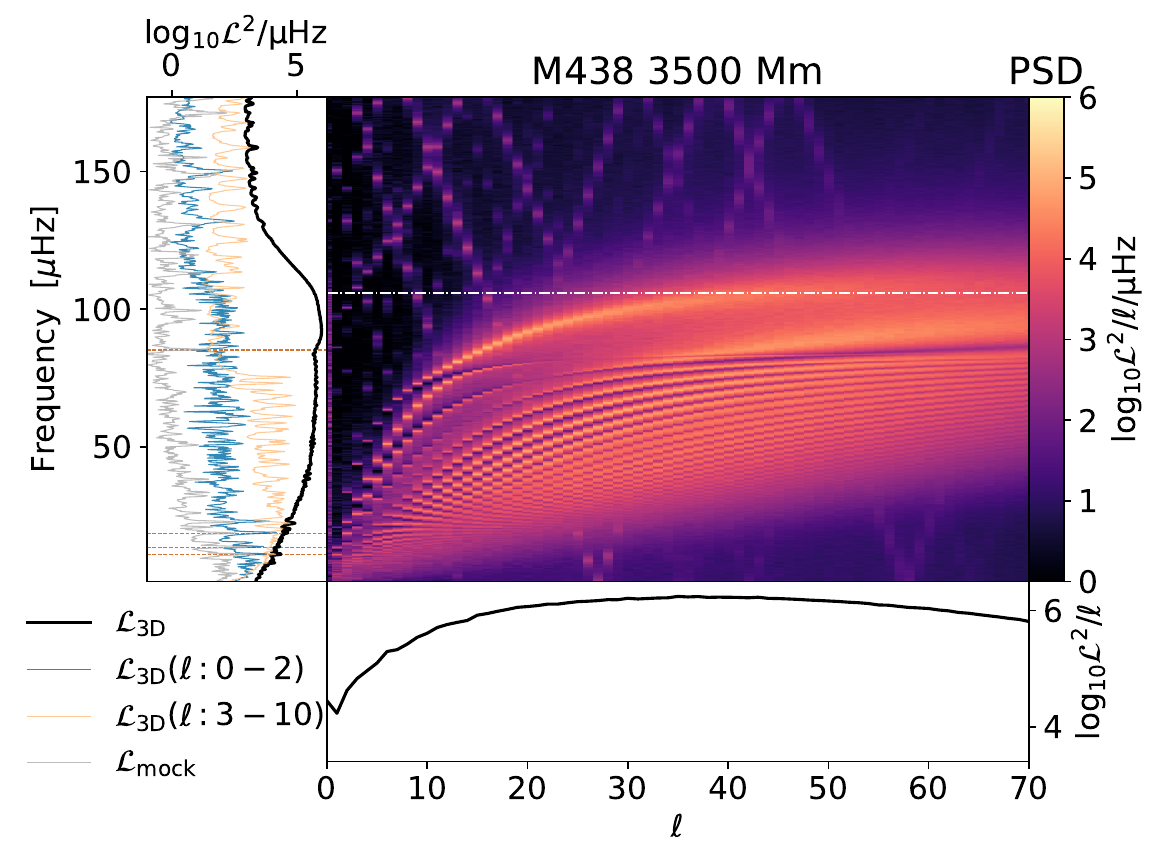}

% Third row
\includegraphics[width=0.49\textwidth]{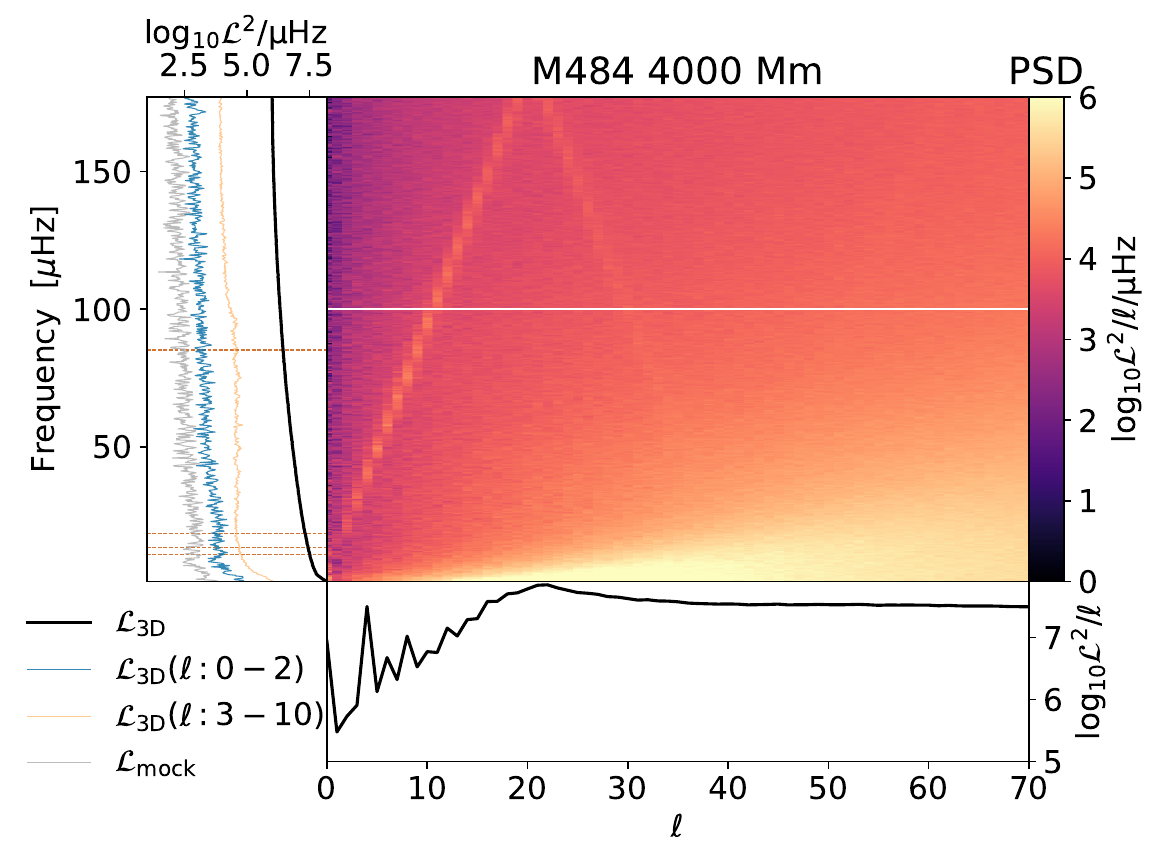}
\includegraphics[width=0.49\textwidth]{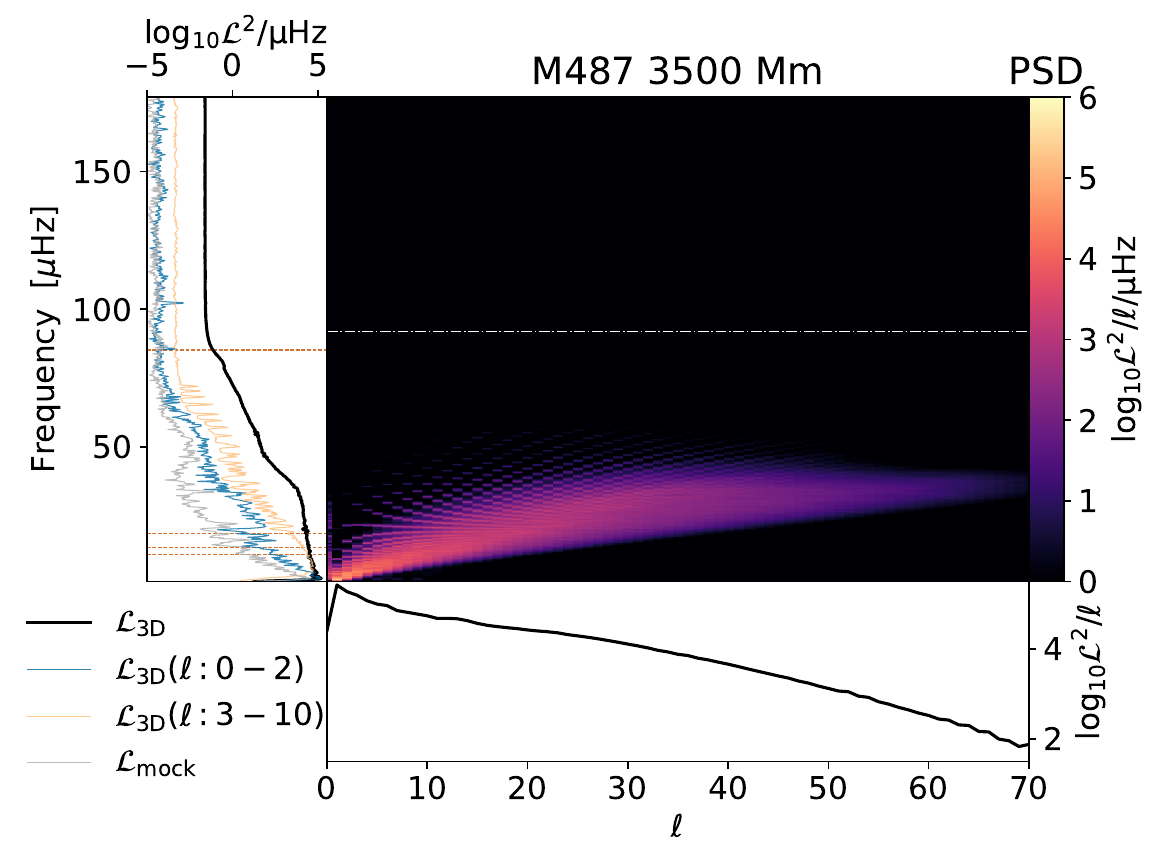}

\caption{Power spectral density (PSD) as a function of spherical harmonic angular degree $\ell$ and cyclic frequency $\nu$ ($\ell-\nu$ diagram) for the variable $\mathcal{L}$ (unity subtracted relative luminosity). Left column shows $\ell-\nu$ diagrams at \unit{4000}{\Mm}, right column at \unit{3500}{\Mm}. \textit{Top row}: Run M424. \textit{Middle row}: Run M438. \textit{Bottom row}: Run M484 (left) and model M487 (right). All $\ell-\nu$ diagrams utilize the final 1600 dumps. Horizontal dashed and dot-dashed lines mark the thin outer envelope convective frequency and linear Brunt--Väisälä frequency $\nu_{\rm Brunt}$. Plots along axes display summed power across each dimension (solid black, \texttt{$\rm \mathcal{L}_{3D}$}). Vertical sub-plot includes power summed over $\ell=0-2$ (blue) and $\ell=3-10$ (orange). Grey curve ($\mathcal{L}_{\rm mock}$) shows luminosity power spectra from Figure~\ref{Fig:power-spectra-4000Mm} and Figure~\ref{Fig:power-spectra-3500Mm} for respective runs.}
\label{fig:komega-diagrams}
\end{figure*}
The $\ell-\nu$ diagram of M424 in Figure~\ref{fig:komega-diagrams} reveals convection spectrum at \unit{4000}{\Mm} (left panel) along with evanescent IGW features. The same evanescent IGW features are present and prominent in the $\ell-\nu$ diagram of M424 at \unit{3500}{\Mm}, within the radiative zone below the thin outer envelope convective boundary, confirming that these features are indeed evanescent IGWs. We see the same IGW features for the run without core heating M438 as in the full-star run M424 with similar spectral power levels in the same Figure~\ref{fig:komega-diagrams}. This implies that these evanescent IGWs are excited at the thin outer envelope convection boundary. The \texttt{$\rm \mathcal{L}_{3D}$} is the power spectral density of 3D luminosity fluctuations around a sphere of fixed radius. $\mathcal{L}_{\rm mock}$ is the power spectral density of line of sight averaged mock luminosity observations obtained by hemispherically integrated 3D luminosity along a line-of-sight. The \texttt{$\rm \mathcal{L}_{3D}$} curve in these $\ell-\nu$ diagrams is a couple of orders of magnitude higher than the grey $\mathcal{L}_{\rm mock}$. This discrepancy comes from the hemispherical integration which involves non-linear cancellation of small scale modes. The difference arises between hemispherically integrated mock luminosity $\mathcal{L}_{\rm mock}$ and the power spectrum of black-body luminosity (Stefan–Boltzmann law) around the sphere calculated using 3D data.

Comparing the $\ell-\nu$ diagrams of the full-star run M424 and the run with core heating turned off M438 reveals the presence of multiple p-modes (above Brunt--Väisälä frequency) and g-modes (below Brunt--Väisälä frequency). Marked peaks in the luminosity power spectrum in Figure~\ref{Fig:power-spectra-4000Mm} of M424 and M438 at \unit{4000}{\Mm} are present both below and above the Brunt--Väisälä frequency, confirming features of both g-modes and p-modes in the spectrum. We also see reflected features of high-frequency p-modes. The dump cadence exceeds the simulation timesteps, causing the Fourier transform algorithm to interpret repeated patterns as modes with frequencies lower than their actual values, which are unresolvable with the dump cadence. The mode features in the $\ell-\nu$ diagrams in Figure~\ref{fig:komega-diagrams} that follow a linear pattern with negative slope represent these reflected high-frequency p-modes, extending below the marked $\nu_{\rm Brunt}$.

\begin{figure*}
\centering
\includegraphics[width=0.8\textwidth]{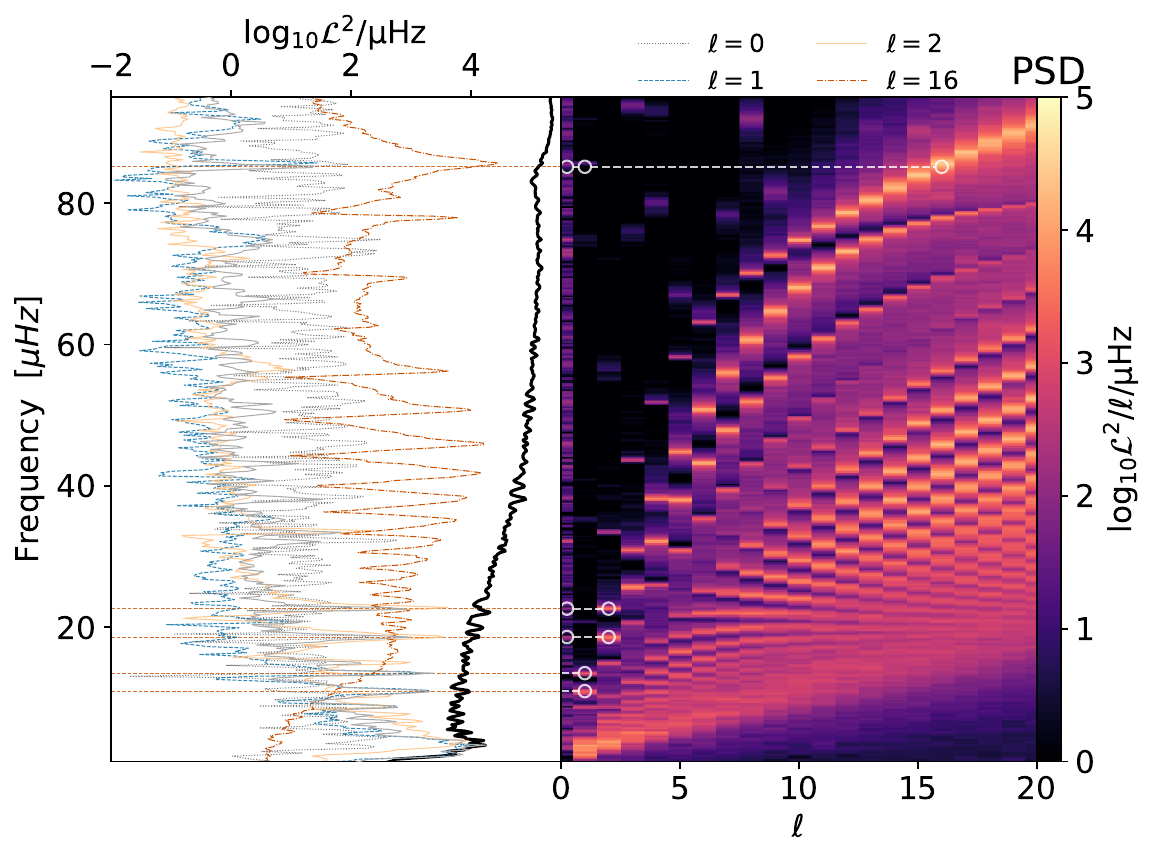}
\includegraphics[width=0.8\textwidth]{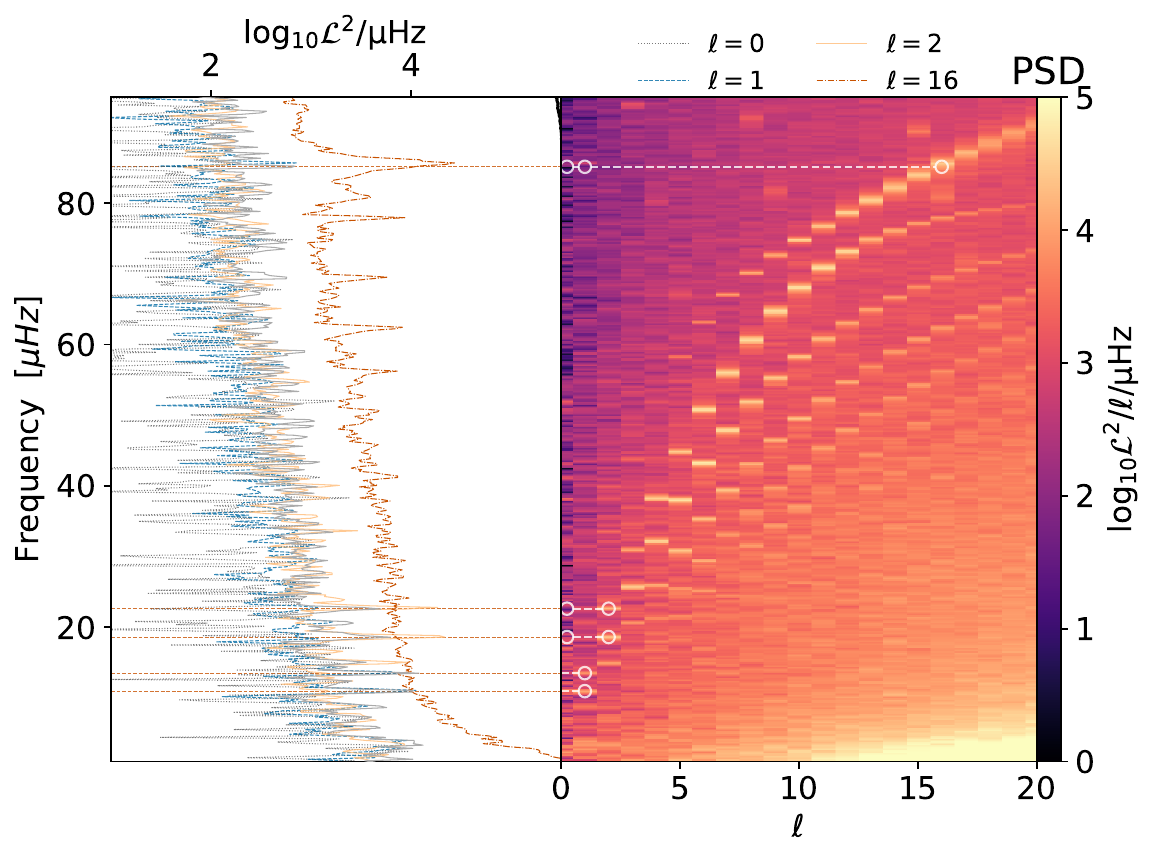}
\caption{Zoomed $\ell-\nu$ diagram for the full-star run M424 at (\textit{Top}) \unit{3500}{\Mm} and (\textit{Bottom}) \unit{4000}{\Mm}, showing $\ell$ range 0--20 and frequency range 0--95 $\mu$Hz. Panel along the y-axis shows \texttt{$\rm \mathcal{L}_{3D}$} and $\mathcal{L}_{\rm mock}$ as in Figure~\ref{fig:komega-diagrams}. Four additional curves show \texttt{$\rm \mathcal{L}_{3D}$} power for $\ell = 0$, 1, 2, and 16. Some features marked in these $\ell-\nu$ diagrams are discussed in the text.}
\label{fig:cropped-komega-M424-3500Mm}
\end{figure*}
The zoomed $\ell-\nu$ diagram for M424 at \unit{3500}{\Mm} in Figure~\ref{fig:cropped-komega-M424-3500Mm} enables identification of spatial scales for the marked features (red vertical dot-dashed lines) in Figure~\ref{Fig:power-spectra-4000Mm} and Figure~\ref{Fig:power-spectra-3500Mm}. The 11 $\mu$Hz and 14 $\mu$Hz features correspond to $\ell = 1$ modes, confirmed by peaks at these frequencies in the $\ell = 1$ curve in Figure~\ref{fig:cropped-komega-M424-3500Mm}. Around 19 $\mu$Hz, we see peaks in both $\ell=0$ and $\ell=2$ curves with power in $\ell=2$ significantly higher than in $\ell=0$ at this frequency. However, since the contribution of these modes to the mock luminosity power $\rm \mathcal{L}_{mock}$ remains unclear, either mode could dominate the observed signal. Near the 85.2 $\mu$Hz feature, we see peaks in $\ell=0$ (85.5 $\mu$Hz) and $\ell=1$ (85.7 $\mu$Hz) p-mode features, and we cannot exclude that this feature could correspond to $\ell=16$ (85.7 $\mu$Hz) g-mode feature which is along the n=-1 arc. However, all the peaks are slightly offset above the feature. Although the spectral power is maximum for the g-mode feature, we still cannot distinguish which mode's contribution is dominant in $\rm \mathcal{L}_{mock}$.

\subsection{Outer convection shell}

The preceding runs M424 and M438 indicate that excitation from core convection has smaller effect to spectral power, suggesting that most low-frequency power originates from the combined effect of IGWs excited near the thin outer envelope convective boundary and from the turbulent thin outer envelope convection zone itself. To quantify the relative contributions of these mechanisms, we conducted run M484, which simulates only the convective envelope.

Figure~\ref{fig:M484-combined} shows the simulated thin convective envelope. For this run, a reflective inner boundary was positioned at $\mathrm{R_{\rm min} = 3850}$ Mm (with $\mathrm{R_{\rm max}=4100}$ Mm, as in M424). For reference, the Schwarzschild boundary of the convective envelope is located at \unit{3813}{\Mm}. This configuration was designed to examine the spectral characteristics of pure convection without IGW interactions. Additional configuration details for simulation M484 are provided in Appendix~\ref{app:M484}.

In Figure~\ref{Fig:power-spectra-4000Mm}, we see that the low-frequency excess power for M484 is significantly reduced compared to M424 and M438 in the envelope region (at \unit{4000}{\Mm}). Total integrated power in M484 is 33\% less as compared to M424. This means that evanescent IGWs in the thin outer envelope convection zone contribute significantly to the low frequency excess along with the turbulent thin outer envelope convection itself. Following the best-fit lime-colored curve to guide the eye about the trend of the spectra of M484, the power spans roughly one-order-of-magnitude from lowest to highest frequencies, qualitatively different from runs M424 and M438 which exhibit a two-order-of-magnitude power difference similar to the power spectrum of observed light curves in Figure~\ref{Fig:observed-spectra}. The shape of the M484 spectrum is flatter and the slope is less variable  compared to the full-star configurations M424 and M438. These runs include a stable zone below the thin outer envelope convection zone that permits the existence and interaction of IGWs with the thin outer envelope convection zone. The characteristic peaks in M424 and M438 spectra are absent in M484 spectra, supporting that g-mode IGWs are absent due to the lack of a stable layer below the convection zone. The $\ell-\nu$ diagram of M484 in Figure \ref{fig:komega-diagrams} (Left) exhibits convection patterns with propagating p-mode features but without evanescent g-mode IGW signatures, consistent with the experimental design. Unlike g-mode IGWs, which require $N^2 > 0$ (stable stratification), p-modes can propagate in regions where $N^2 < 0$, the condition in convection zones.

\subsection{No outer convection (large radiative envelope)}

Simulation M487 is like M424 but excludes the thin outer envelope convection zone to isolate the effect of core convection on the low frequency excess. Due to the radiation diffusion timestep constraints discussed in Section~\ref{sec:Methods}, it is impossible with our approach to execute a simulation without the envelope convection zone while maintaining $\mathrm{R_{\rm max} = 4100~\rm{Mm}}$. Consequently, for simulation M487, we positioned the outer boundary $\mathrm{R_{\rm max}}$ radially inward at $3600~\rm{Mm}$ (inner 83\% by radius of the ZAMS MESA model). As a result, run M487 would have different eigenfrequencies compared to the full-star run M424 because its stratification is different.

\begin{figure}
   \centering
   \begin{interactive}{animation}{animations/M487-merged_movie.mp4}
   \includegraphics[width = \hsize]{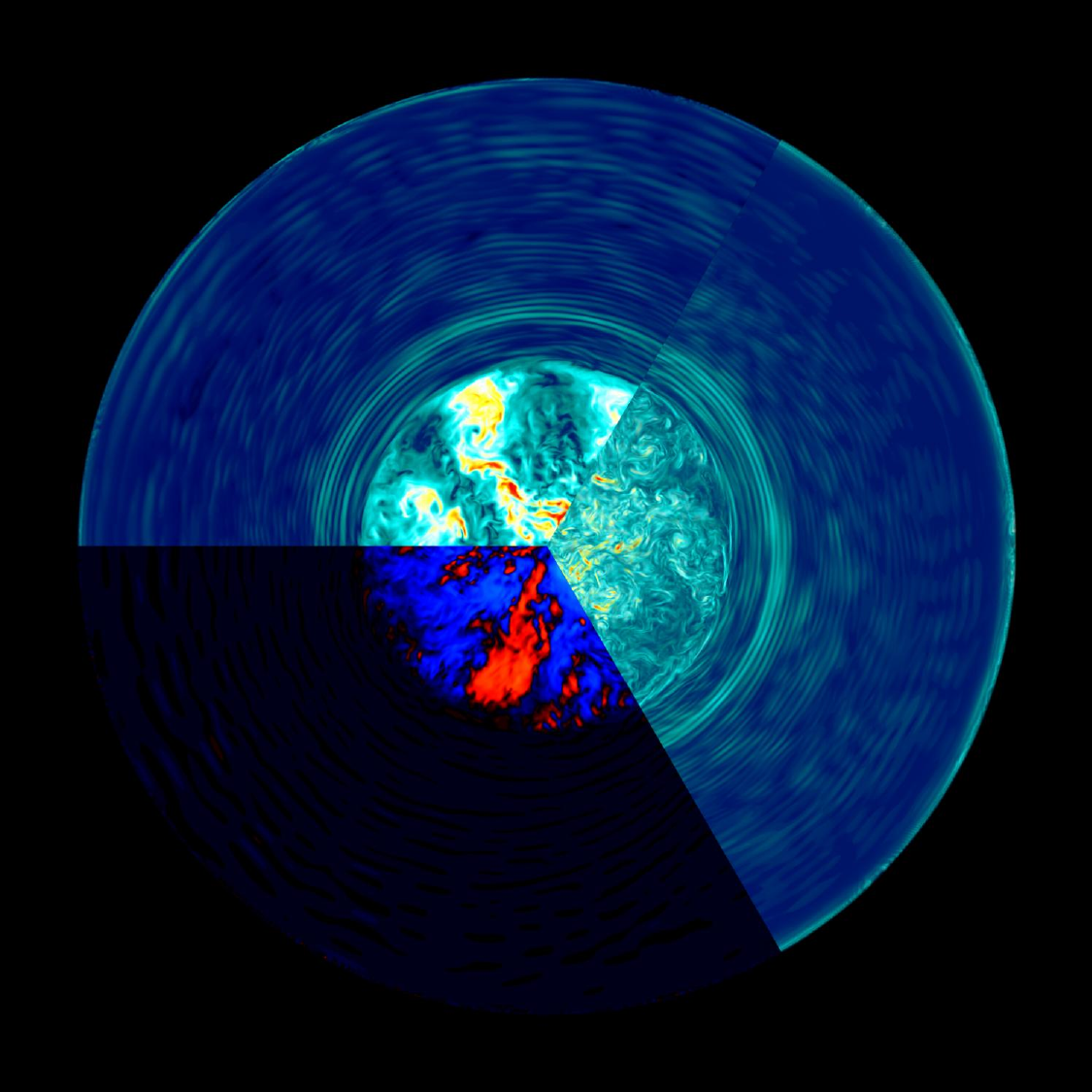}
   \end{interactive}
   \caption{Volume rendered images similar to Figure~\ref{Fig:M424-combined} but for the run M487. An animation is available in the HTML version showing the temporal evolution over 30 consecutive dumps.}
   \label{fig:M487-combined}
\end{figure}
As shown in the bottom panel of Figure~\ref{Fig:power-spectra-3500Mm}, M487 exhibits comparable or greater low-frequency excess levels relative to M424 at \unit{4000}{\Mm}. At frequencies $\mathrm{\leq\; 10\mu Hz}$, the power levels of M487 and M424 at \unit{3500}{\Mm} are approximately equivalent, whereas at frequencies $\mathrm{>\; 10\mu Hz}$, M487 demonstrates reduced power levels compared to M424. These results indicate that at \unit{3500}{\Mm}, the lower frequency regime ($\mathrm{\leq\; 10\mu Hz}$) is predominantly influenced by IGWs originating from the core, while the higher frequency component is primarily attributable to the presence of the thin outer envelope convection zone. This is further supported by the $\ell-\nu$ diagram for M487 presented in the bottom right panel of Figure \ref{fig:komega-diagrams}, which shows that the maximum power is concentrated at the lowest frequencies and largest spatial scales (small $\ell$ values). 

The $\ell-\nu$ diagram in Figure~\ref{fig:komega-diagrams} (bottom right) for run M487 appears significantly different from the $\ell-\nu$ diagrams presented in \citet{Mao2024} for simulations without thin outer envelope convection zones. The $\ell-\nu$ diagram of the run M487 at \unit{3500}{\Mm} shows significantly less power in the high frequency ($\nu\gtrsim 40\, \rm\mu Hz $) eigenmodes as compared to \citet{Mao2024}. The low radial order eigenmodes ($n=-1,-2,-3$) that are clearly visible in the $\ell-\nu$ diagrams of \citet{Mao2024} have much lower power levels in run M487, making them not visible in the $\ell-\nu$ diagram (Figure~\ref{fig:komega-diagrams}). This leads us to conduct another simulation.

\subsection{No outer convection (small radiative envelope)}
\begin{figure}
\begin{interactive}{animation}{animations/M488-merged_movie.mp4}
   \centering
   \includegraphics[width = \hsize]{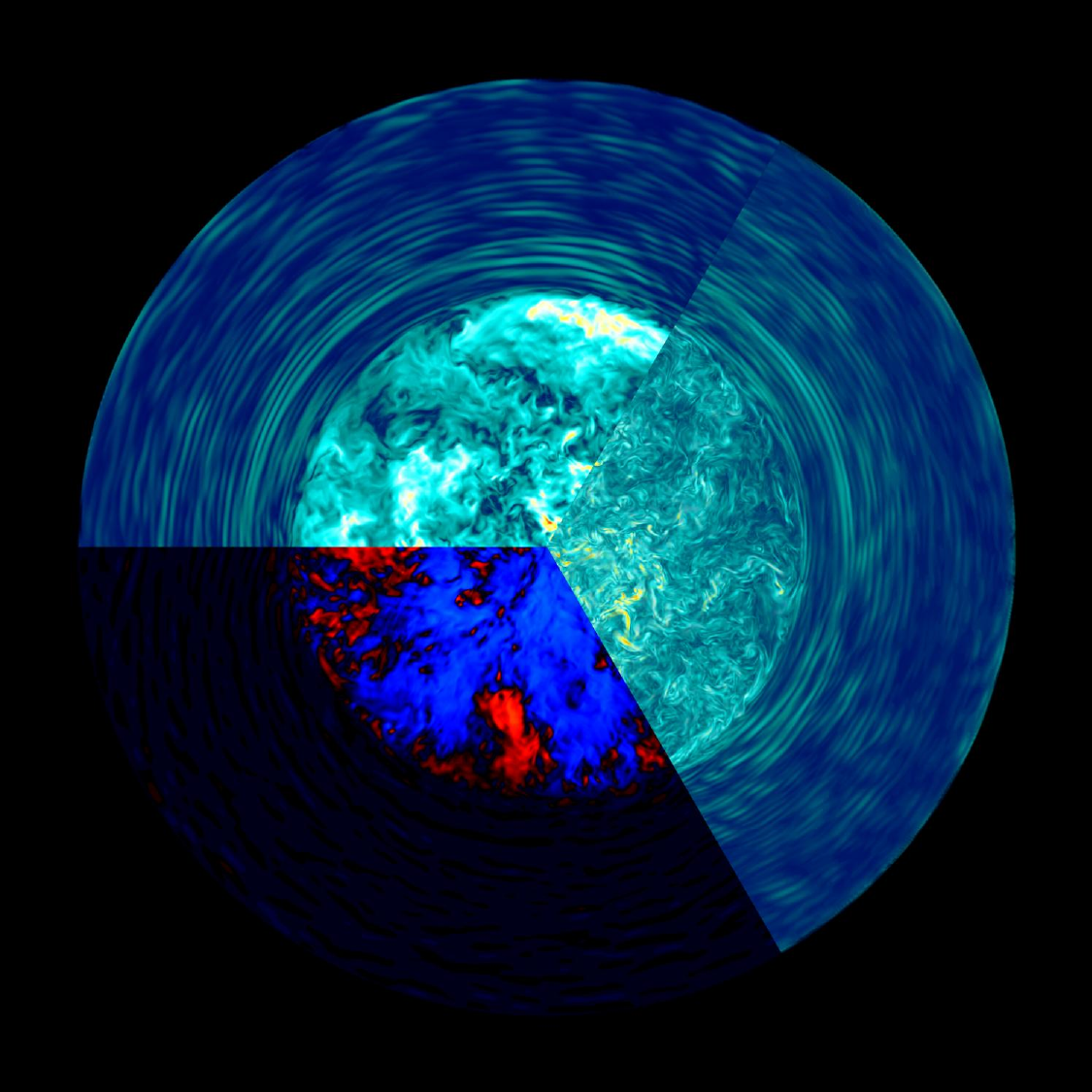}
\end{interactive}
   \caption{Volume rendered images similar to Figure~\ref{Fig:M424-combined} but for the run M488. An animation is available in the HTML version showing the temporal evolution over 30 consecutive dumps.}
   \label{fig:M488-combined}
\end{figure}
 To determine whether this difference results from the additional \unit{930}{\Mm} radiative envelope in M487 which has larger radiative damping as compared to the runs in \citet{Mao2024}, we conducted run M488. This simulation is identical to M487 but with the outer boundary $\rm R_{max}$ positioned at \unit{2670}{\Mm}, matching the configuration in \citet{Mao2024}, as shown in Figure~\ref{fig:M488-combined}.
\begin{figure*}
   \centering
   \includegraphics[width = 0.495\textwidth]{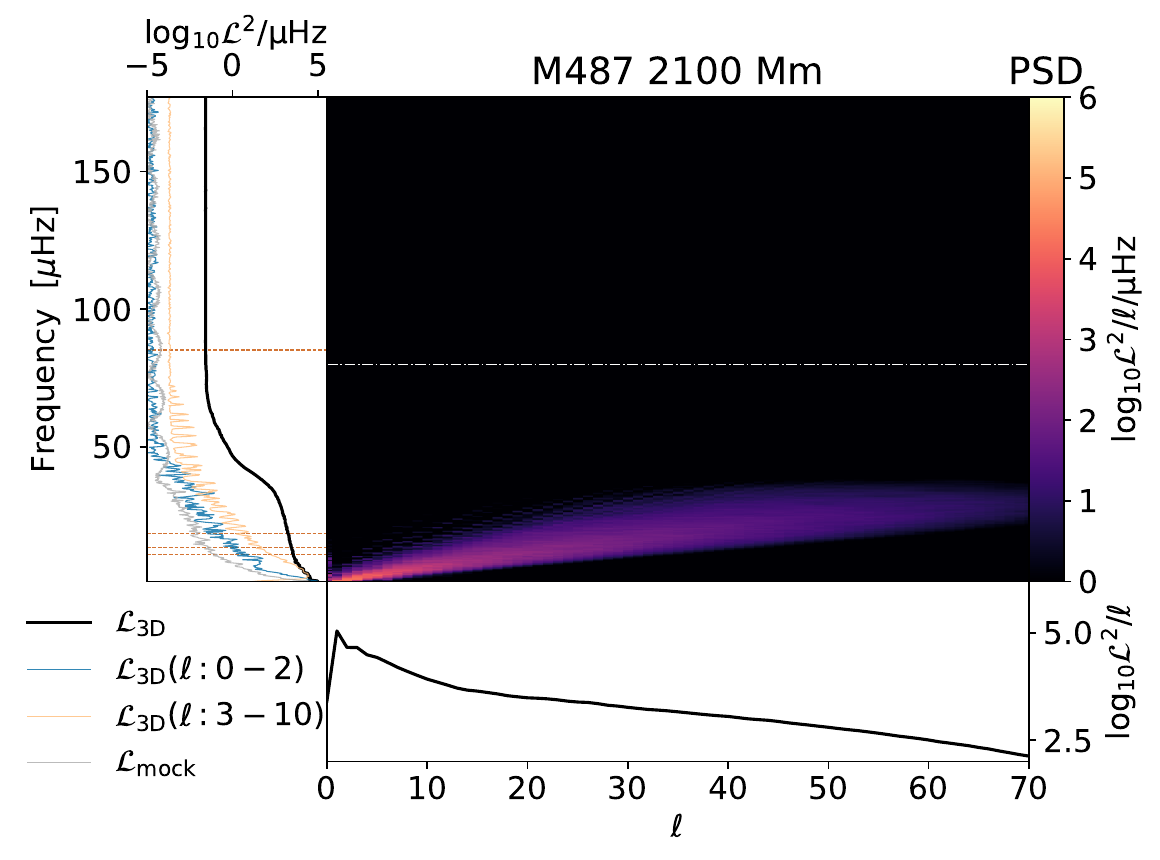}
   \includegraphics[width= 0.495 \textwidth]{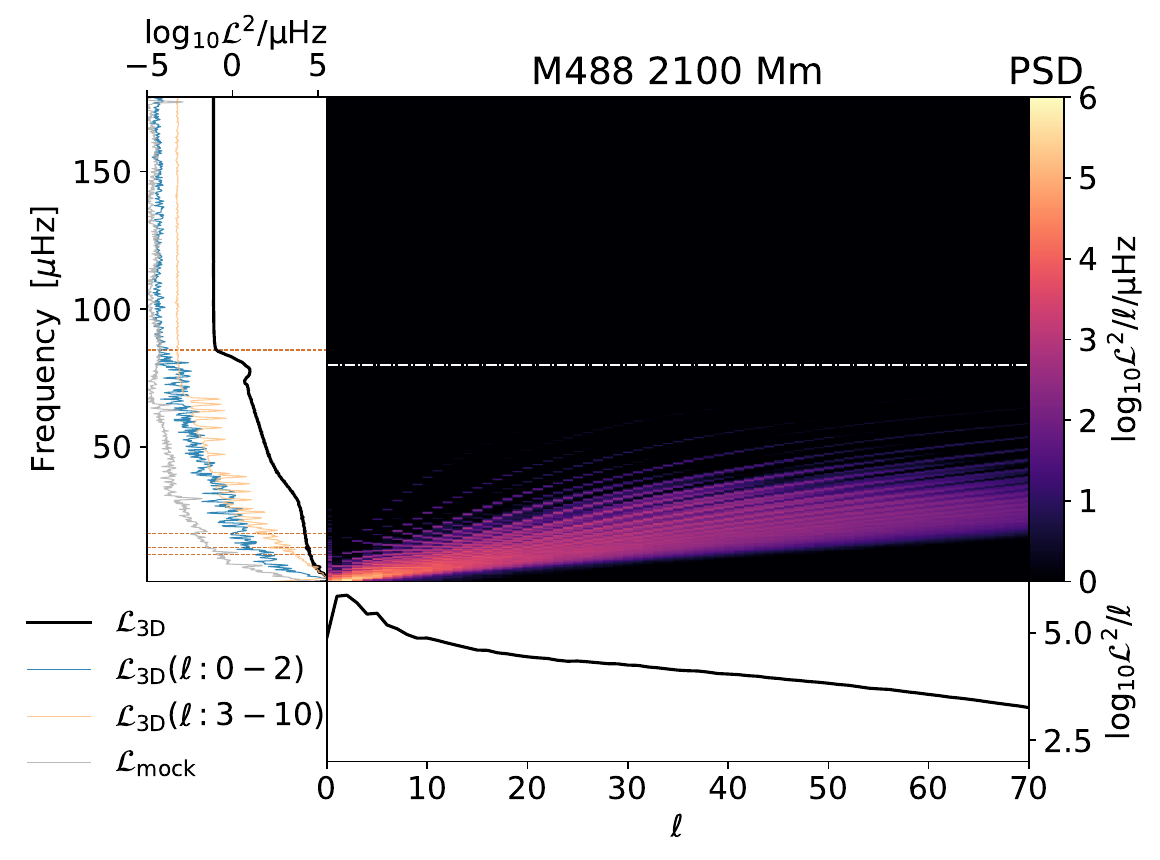}
   \caption{ The $\ell-\nu$ diagram for the run M487 and M488 at \unit{2100}{\Mm} similar to Figure~\ref{fig:komega-diagrams}.}
   \label{komegaM488-M487_2100Mm}
\end{figure*}

The $\ell-\nu$ diagram of run M488 in Figure~\ref{komegaM488-M487_2100Mm} exhibits similar characteristics to M487 at \unit{2100}{\Mm}. We see the suppression of higher frequency ($\nu \gtrsim 40\, \mu$Hz) eigenmodes in the $\ell-\nu$ diagrams of these runs. However, there are differences between the $\ell-\nu$ diagrams of M487 and M488 in Figure~\ref{komegaM488-M487_2100Mm}. Run M487 with the large envelope shows dominantly traveling waves appearing as diagonal ridges or tracks that extend continuously across multiple $\ell$ values, whereas M488 shows distinct eigen modes with little evidence of traveling waves. These results are discussed in the following Section~\ref{sec:discussion-and-conclusion}.

\subsection{Wave excitation and  damping}

\begin{figure*}
   \centering
   \includegraphics[width = 0.495\textwidth]{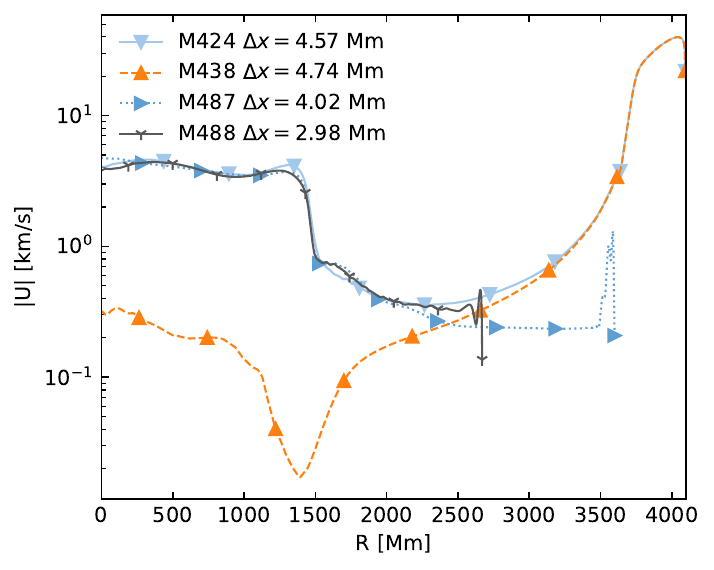}
   \includegraphics[width= 0.495 \textwidth]{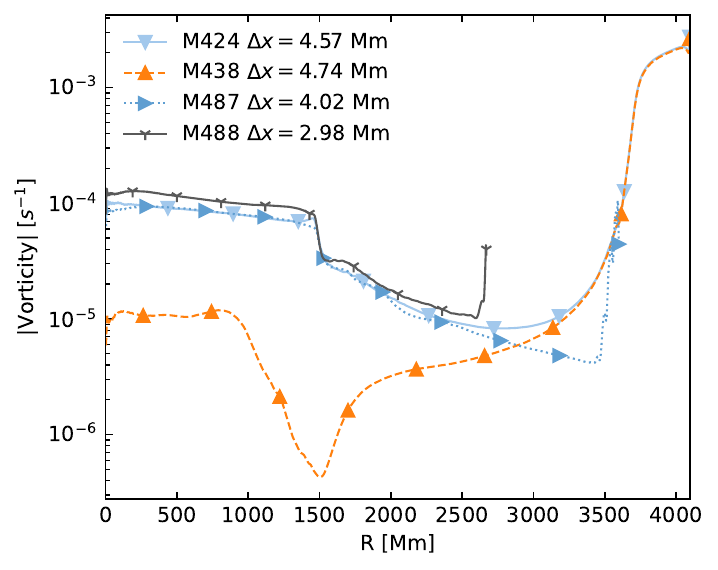}
   \caption{Radial profile of (\textit{Left}) total velocity magnitude $|U|$ and (\textit{Right}) vorticity magnitude averaged over last 400 dumps of respective runs. }
   \label{Fig:U_vorticity_rad_profile}
\end{figure*}
Figure~\ref{Fig:U_vorticity_rad_profile} shows the radial profiles of velocity
magnitude and vorticity for runs with radiative zones (M424, M438, M487, M488). The left
panel shows that in runs with core convection waves have high amplitudes in the inner radiative region ($R < 2500$
Mm) while simulations with outer convection zone feature high wave amplitudes in the outer radiative region ($R > 3000$
Mm). The intermediate region around \unit{2600}{\Mm} shows in run M424 the combined excitation from both convection zones.

In simulations like these, ideally numerical viscosity is smaller than radiative diffusion. To get insight on this we consider vorticity magnitude, which is sensitive to contributions from
small scales. In the $\unit{25}{\Msun}$ main-sequence star simulations without radiation, \citet[][Figure 29]{Herwig2023} vorticity magnitude in the radiative envelope increases with finer grids, across four grid resolutions covering a factor $3.5$. This demonstrates that as the grid resolution increases smaller scales in the wave region are resolved and contribute to the vorticity magnitude, just as in the convective core.  In contrast, \citet[][Figure 15]{Mao2024} shows identical vorticity levels 
in the wave region ($r > \unit{1604}{\Mm}$) across the same four grid resolutions in simulations with radiative diffusion. This demonstrates that in those simulations where both the luminosity and the radiative diffusion is boosted by a factor $1000$,  radiative diffusion dissipates small scales that are larger than what can be resolved by the lowest-resolution grid shown ($768^3$,$\Delta x = \unit{6.5}{\Mm}$), and that radiative diffusion dominates over numerical dissipation in those simulations. In these simulations the grid resolution is higher ($\Delta x \approx 4.0$ to $\unit{4.6}{\Mm}$) but the thermal diffusivity is 10 times smaller (boost factor 100). Still, runs M424 and M488 with a resoluion difference of 1.5  exhibit identical
vorticity levels for wave motions, demonstrating that also in these 100x boost simulations with the grids deployed here, thermal diffusion dominates over numerical viscosity.

The  wave power spectrum in the simulations results from nonlinear wave excitation and de-excitation
at the convective boundaries combined with frequency-dependent radiative damping,
including in run M487 at \unit{3500}{\Mm} (Figure~\ref{Fig:power-spectra-3500Mm}) where
the power contrast between low and high frequencies is particularly pronounced. Wave
excitation is a nonlinear process where turbulent convection generates waves with
amplitudes determined by the convective properties. The wave excitation is largest at low
frequencies near the core convective frequency of $\approx \unit{1.27}{\mu \mathrm{Hz}}$
\citep[computed as $U_{\mathrm{\rm core}}/2R_{\mathrm{\rm core}}$ where $R_{\mathrm{\rm
core}}$ is the convective core radius and $U_{\mathrm{\rm core}}$ is the average core
convective velocity,][]{Thompson2024}.

\begin{figure*}
   \centering
   \includegraphics[width = 0.495\textwidth]{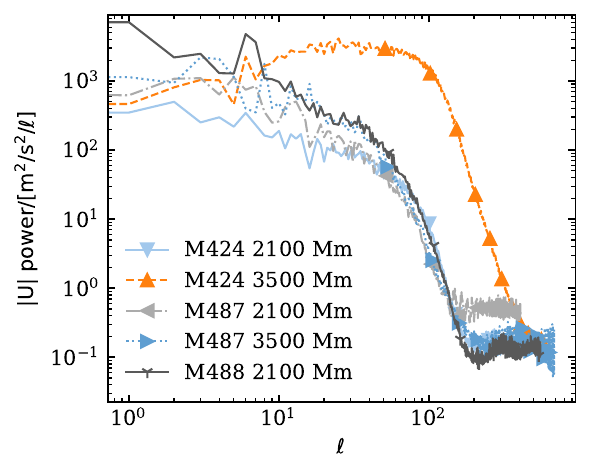}
   \includegraphics[width= 0.495 \textwidth]{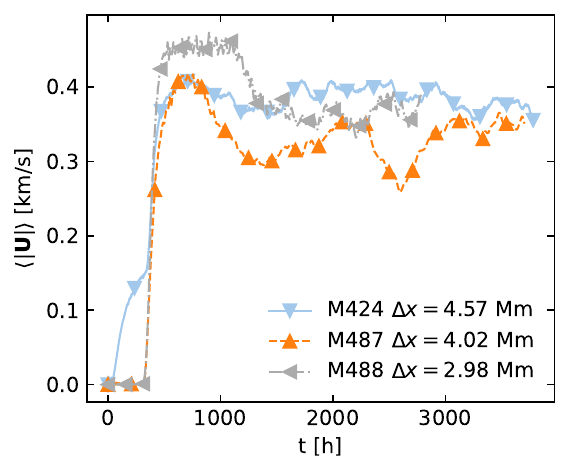}
   \caption{\textit{Left}: Spatial spherical harmonic spectra of total velocity magnitude $|U|$ for the runs with core heating at \unit{2100}{\Mm} and \unit{3500}{\Mm} inside the radiative zone. \textit{Right}: Time series of $|U|$ averaged over \unit{2000}{\Mm} to \unit{2200}{\Mm} for the runs with core heating. }
   \label{ell_spectra_and_U_av}
\end{figure*}

Previous \texttt{PPMstar} simulations of a \unit{25}{\mathrm{M_\odot}} mid-main-sequence star without the outer convection zone demonstrate that low-frequency waves experience less radiative damping than high-frequency waves as compared to a reference simulation without radiative damping \citep[][Figure 14]{Mao2024}. Figure~\ref{ell_spectra_and_U_av} (Left) shows that numerical dissipation becomes significant only at spatial scales $\ell > 100$, well above the range of $\ell \leq 70$ examined in our $\ell-\nu$ diagrams (Figure~\ref{fig:komega-diagrams}). Similar behavior is observed in  \citet[][Figure 18]{Thompson2024} across different resolutions and especially for run M114 with grid resolution $\Delta x = 4.69$ Mm, comparable to our $1792^3$ full-star run M424 with $\Delta x = 4.58$ Mm and $1792^3$ run M487 with $\Delta x = 4.02$ Mm. Figure~\ref{ell_spectra_and_U_av} (Right) shows that wave velocities in the radiative zone are consistent across resolutions different by by a factor 1.5. A wider range grid resolutions covering a factor 3.5 reported in Figure~8 (bottom panel) in \cite{Mao2024}  shows that wave amplitudes do not depend on resolution. The evidence from vorticity (Figure~15) has already been discussed earlier in this section.   Furthermore, the frequencies and wavenumbers of eigenmodes in our simulations are consistent with linear theory predictions from \texttt{GYRE} \citep[][Figure~13]{Thompson2024}. And finally, \texttt{PPMstar} simulations of red giant branch star \citep[][Figure 14]{Blouin2023} show that IGW damping in the radiative zone agrees with linear theory \citep{Zahn1997}.

\section{Discussion and Conclusion}
\label{sec:discussion-and-conclusion}

We conducted 3D hydrodynamic simulations of a non-rotating 25 $\mathrm{M_\odot}$ ZAMS star, encompassing approximately 95\% of the stellar structure and including both core and thin outer envelope convection zones. Analysis of the mock lightcurve power spectra reveals three key findings. The luminosity power spectra for the full-star run (M424) near the envelope exhibit morphological characteristics qualitatively and quantitatively similar to the observed SLF spectra. Comparison of spectra between the full-star run (M424) and the no-core-heating run (M438) at \unit{4000}{\Mm} demonstrates small (10\%) contribution from the convective core and IGWs excited at the convective core boundary to the low-frequency excess. However, this reduced contribution to observable power does not imply that observable signatures cannot probe core stellar structure. The IGW features present in the mock luminosity spectra depend on the entire stellar stratification, including the core structure, which could enable inference of interior properties from surface observations. Comparative analysis of the full-star run (M424) and the envelope-only run (M484) at \unit{4000}{\Mm} indicates that the low-frequency excess results from combined effects of IGWs and pure thin outer envelope convection, with IGWs being the dominant contributor to excess power at low frequencies, as evidenced by the $\mathrm{a_0 + C_w}$ values near $\nu = 0$.

The $\ell-\nu$ diagrams in Figure \ref{fig:komega-diagrams} for all simulations within the thin outer envelope convection zone at \unit{4000}{\Mm} show maximum power spectral density at frequencies below $10~\mu$Hz. This frequency distribution appears in the mock luminosity power spectra in Figure~\ref{Fig:power-spectra-4000Mm}, where dominant power occurs at frequencies $< 10~\mu$Hz across all simulations at \unit{4000}{\Mm} (left column). The thin outer envelope convection turnover timescale is $\approx 3~$h, corresponding to a convective frequency of $96~\mu$Hz. Given a convective frequency, one would expect the maximum power spectral density to occur around that frequency for a convection-dominated region, but the 3D simulations show dominant power at frequencies of order unity $\mu$Hz, almost an order of magnitude lower than the envelope convective frequency. This pattern was also observed in the 3D wedge simulations of a non-rotating thin outer envelope convection zone of a $35 \mathrm{M_\odot}$ main sequence star by \citet{Schultz2022}. Our estimate of their convective frequency is $13~\mu$Hz\footnote{Convective frequency was not directly reported in \citet{Schultz2022}. We used the ratio of convective velocity and the radial extent of convection zone to determine the convective frequency.}, while their Lorentzian characteristic frequency $\nu_{\mathrm{char}}\approx 110 \, \mu$Hz. Despite this characteristic frequency being larger than their convective frequency, their simulations also showed the majority of power spectral density at much lower frequencies, similar to our findings, although they did not explicitly emphasize this point.

This discrepancy between simple convective frequency estimates and observed spectral distributions indicates a fundamental issue with assuming circular convective cells. Figure~\ref{Fig:Spatial_spectra-M424} shows the dominant spatial scale of $\ell\approx 40$, which corresponds to the radial extent of the thin outer envelope convection zone. Given the velocities shown in Figure~\ref{Fig:FeCZ-convective-velocities-M424}, circular convective cells would require sizes 10 times larger to accommodate the observed lower frequencies, but this is inconsistent with the observed spatial scales. The convective cells cannot be elongated horizontally either, as this would appear in the tangential velocity spatial spectrum.

The $\ell-\nu$ diagram of the global luminosity fluctuations $\mathcal{L}_{\rm 3D}$ in Figure~\ref{fig:komega-diagrams} show higher spectral power density in run M438 compared to the full-star run M424. This appears counterintuitive because M438 omits core volume heating. Core convection is the primary driver of internally generated gravity waves (IGWs) at the stellar core. However, the core remains isentropic in M438, causing it to function as a nearly perfect reflector of IGWs. The core in M424 acts as an imperfect reflector due to turbulent convective motions. The net power in the global luminosity fluctuations $\mathcal{L}_{\rm 3D}$ of the envelope convection run M484 shows higher spectral power density than the full-star run M424. This enhancement results from two nearby reflecting boundaries in M484, which produce elevated flow velocities.

Our simulations differ from previous 3D hydrodynamic simulation studies of massive main sequence stars addressing SLF variability. Previous studies simulated a wedge or partial stellar structure \citep[e.g.][]{Schultz2022,Schultz2023}. We simulate the star with both core and envelope convection zones in 3D and full $4\pi$ geometry. This approach resolves modes and eigenwaves at the scale of the entire stratification and assesses their impact on the power spectrum. We take timesteps that are limited by sound speed according to the CFL condition \citep{Courant1928}, thereby resolving pressure waves and their interactions and effects in the dynamics, as opposed to simulations with timesteps limited by the flow speed, and hence orders of magnitude larger timesteps, like \citet{Anders2023}.

The $\ell-\nu$ diagrams at \unit{2100}{\Mm} for the runs M487 and M488, without the thin outer envelope convection, look qualitatively different from the $\ell-\nu$ diagrams presented in \citet{Mao2024}. We see the suppression of high frequency ($\nu \gtrsim 40\, \mu$Hz) eigenmodes in the $\ell-\nu$ diagrams of runs M487 and M488 at \unit{2100}{\Mm} (Figure~\ref{komegaM488-M487_2100Mm}) compared to the $\ell-\nu$ diagrams of the runs in \citet{Mao2024}, where we can clearly see power in low radial order eigenmodes ($n=-1,-2,-3$) up to the maximum Brunt--Väisälä frequency of the envelope. Run M487 with the large envelope shows dominantly traveling waves, whereas M488 shows distinct eigen modes and little evidence of traveling waves. The probable cause is that in M487 with the stable layer extending to much larger radii, radiative diffusion damping prevents waves from being effectively reflected. This would also be true in the full-star run M424, supporting the argument that waves excited by core convection do not reach the outer convection zone with significant power and contribute less (10\%) to the observable temporal spectrum.

% This could result from two possible differences between our runs and those in \citet{Mao2024}. First, we boost the core heating luminosity and radiative diffusivity by a factor of 100 compared to a factor of 1000 in \citet{Mao2024}. Second, there is a mean molecular weight ($\mu$) gradient in the runs of \citet{Mao2024} unlike our ZAMS runs with constant $\mu$. The $\mu$ gradients may be essential in exciting waves because they change the boundary stiffness in ways that are not fully understood. This would mean that the ZAMS setup is different or special, in that its core convection does not excite waves the way later phase simulations do, due to differences in chemical stratification. If this is true, then our findings apply only to ZAMS stars, and not generally to all main-sequence stars. We are planning to investigate this in future.

Several limitations must be acknowledged when interpreting these results. Our simulations were performed with both core luminosity and radiative diffusivity set to 100 times their nominal values from the \texttt{MESA} model. It has been observed that core convective velocities scale with core luminosity boost factor as (boost factor)$^{-1/3}$ \citep{Porter2000,Jones2017,Herwig2023}. This means that our core convective velocities are enhanced by a factor of approximately 4.64, while the core convective turnover timescale is reduced by the same factor compared to the \texttt{MESA} model. 

\begin{figure}
   \centering
   \includegraphics[width=\hsize]{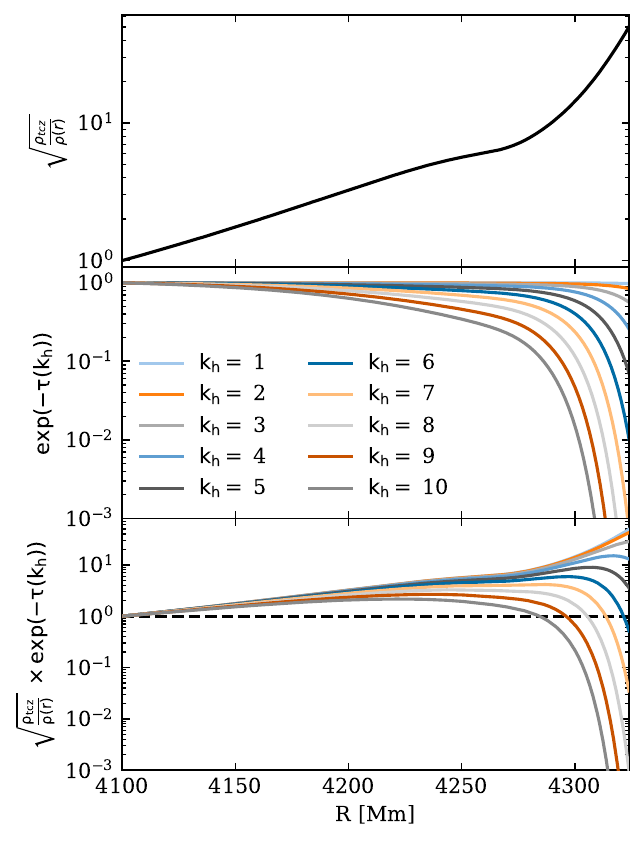}
      \caption{Radial dependence of IGW amplification and damping factors relative to the wave amplitude at \unit{4100}{\Mm}. \textit{Top panel:} Wave amplification factor $\sqrt{\rho_{\mathrm{tcz}}/\rho(r)}$ due to decreasing density with radius, applicable to all horizontal wavenumbers $k_h$. \textit{Middle panel:} Wave damping factor $\exp[-\tau(\omega, k_h, r)]$ for horizontal wavenumbers $k_h = 1$ to $10$. \textit{Bottom panel:} Combined amplification and damping factor $\nu_{\mathrm{wave}}(\omega, k_h, r)/\nu_{\mathrm{rms-cz}}$ for $k_h = 1$ to $10$. The curves extend from the outer boundary of the full-star simulation at \unit{4100}{\Mm} to the outer boundary of the \texttt{MESA} model at \unit{4322}{\Mm}. Labels are shown only in the middle panel.}
         \label{Fig:damp-amplify-igw}
\end{figure}

Our simulations exclude the outer 5\% of the stellar radius. Waves propagating toward the outer regions could experience additional damping due to radiative diffusion 
\citep{Buehler2009} 
and amplification due to decreasing density \citep[pseudomomentum conservation,][]{Kumar1999}. Following \citet{Rogers2017}, the wave amplitude near the surface, accounting for both effects, can be estimated as:
\begin{gather}
    \nu_{\mathrm{wave}} (\omega,k_h,r) = \nu_{\mathrm{rms-cz}} \sqrt{\frac{\rho_{\mathrm{tcz}}}{\rho(r)}}\, \exp{[-\tau (\omega, k_h,r)]} \\
    \implies \frac{\nu_{\mathrm{wave}} (\omega,k_h,r)}{\nu_{\mathrm{rms-cz}}} = \sqrt{\frac{\rho_{\mathrm{tcz}}}{\rho(r)}}\, \exp{[-\tau (\omega, k_h,r)]} \label{eq:wave-damp/amplification}
\end{gather}
where $\nu_{\mathrm{wave}} (\omega,k_h,r)$ represents the wave amplitude at outer radial location $r$ for a wave of frequency $\omega$, $\nu_{\mathrm{rms-cz}}$ is the rms velocity within the convection zone, $k_h$ is the horizontal wavenumber, $\rho_{\mathrm{tcz}}$ is the density at the top of the convection zone, $\rho(r)$ is the density at the outer radial location, and $\tau (\omega, k_h,r)$ is the wave damping optical depth defined as \citep{Kumar1999}:
\begin{equation}
    \tau (\omega, k_h,r) = \int_{r_{\mathrm{tcz}}}^r \frac{\kappa k_h^3 N^3}{\omega^4 (2\pi r)^3} \, dr
\end{equation}
where $r_{\mathrm{tcz}}$ is the radius at the top of the convection zone, $\kappa$ is the radiative diffusivity at radial location $r$.

Using parameters from the ZAMS \texttt{MESA} model, maximum Brunt--Väisälä frequency near the surface $N \rm\approx 2\pi\times270\, rad\,s^{-1}$, frequency of the sharpest observed peak in M424 luminosity power spectrum in Figure~\ref{Fig:power-spectra-4000Mm} $\omega = 2\pi \times 10\,\mu$Hz, and assuming the thin outer envelope convection zone terminates at \unit{4100}{\Mm} with a radiative zone extending to the outer boundary in the \texttt{MESA} model at \unit{4322}{\Mm}, we calculate relative wave amplification and dissipation factor radial profiles using Equation~\ref{eq:wave-damp/amplification}. Figure~\ref{Fig:damp-amplify-igw} shows that amplitudes of IGWs with $k_h \leq 5$ are amplified whereas higher $k_h$ amplitudes are damped. The $\ell-\nu$ diagrams in Figure~\ref{fig:komega-diagrams} indicate that the peak at around $10\,\mu$Hz corresponds to $\ell \leq 5$. Therefore, assuming the extrapolation using Equation~\ref{eq:wave-damp/amplification}, some of the marked IGW peaks in Figure~\ref{Fig:power-spectra-4000Mm} could be visible near the photosphere.

The implementation of a reflecting outer boundary at \unit{4100}{\Mm}, necessitated by challenges in resolving radiation diffusion in the outermost layers, could artificially modify standing IGW amplitudes. This may influence the power spectrum near the surface, as IGW features are evident in the spectrum. This artificial effect is inherent to our simulation approach. Quantifying its impact would be valuable, though the methodology for doing so remains unclear.

\section*{acknowledgments}
We thank Dominic Bowman for sharing the TESS and CoRoT timeseries data for the star \texttt{HD46150} and meaningful discussions. We thank Daniel Lecoanet for valuable discussions and comments on the manuscript. P.R.W. acknowledges support from NSF CDS\&E grants 1814181 and 2309101. The simulations for this work were carried out on the NSF Frontera supercomputer operated by the Texas Advanced Computing Center at the University of Texas at Austin and on Compute Canada's Niagara supercomputer operated by SciNet at the University of Toronto. We thank Ted Wetherbee whose scripts were used to generate volume rendered images. The data analysis was carried out on the Astrohub virtual research environment (\url{https://astrohub.uvic.ca/}) hosted on the Digital Research Alliance of Canada Arbutus Cloud at the University of Victoria. LLMs were used to improve wording at the sentence level and assist with coding.

\vspace{5mm}
\software{\texttt{PPMstar}, \texttt{MESA}}

\bibliography{references}{}
\bibliographystyle{aasjournal}

\appendix
\FloatBarrier
\renewcommand{\thefigure}{A\arabic{figure}}
\renewcommand{\thetable}{A\arabic{table}}
\setcounter{figure}{0}
\setcounter{table}{0}

\begin{figure*}
\includegraphics[width=0.49\textwidth]{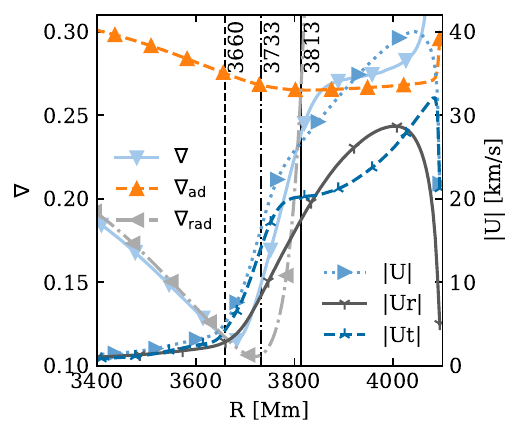}
\includegraphics[width=0.49\textwidth]{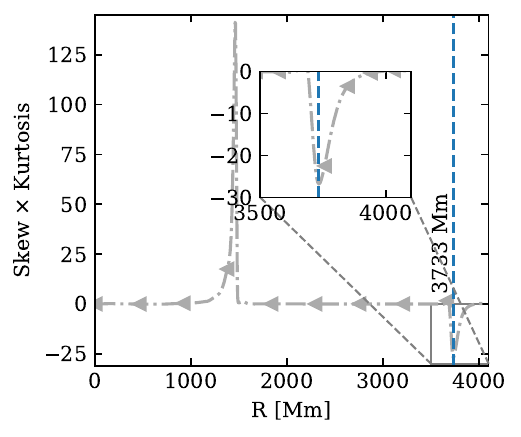}
\caption{\textit{Left}: Radial profiles of the actual temperature gradient ($\nabla$; light blue solid line with downward triangles), adiabatic temperature gradient ($\mathrm{\nabla_{ad}}$; orange dashed line with upward triangles) and radiative temperature gradient ($\mathrm{\nabla_{rad}}$; grey dot-dashed line with leftward triangles) at dump 4800 ($\mathrm{\approx 4779\,h}$) for run M424 near the thin outer envelope convection zone. The secondary y-axis displays total velocity magnitude $\mathrm{(|U|)}$, radial velocity magnitude $\mathrm{(|U_r|)}$ and tangential velocity magnitude $\mathrm{(|U_t|)}$ at dump 4800. \textit{Right}: Radial profile of $\mathrm{Skew \times Kurtosis}$ of radial velocity $\mathrm{U_r}$ for run M424. The vertical dashed dark blue line marks the extremum at \unit{3733}{\Mm}, with an expanded view of this region shown in the inset.}
\label{fig:conv-boundary-M424}
\end{figure*}
\section{Outer Envelope Convective Boundary}\label{app:convective-boundary}
We analyze the thin outer envelope convective boundary and overshoot region for run M424 using temperature gradient profiles and statistical moments of the radial velocity field. The left panel of Figure~\ref{fig:conv-boundary-M424} shows that from the interior toward the surface, the temperature gradient $\nabla$ follows the radiative gradient $\mathrm{\nabla_{rad}}$ up to \unit{3660}{\Mm}. Beyond this point, $\nabla$ increases toward $\mathrm{\nabla_{ad}}$ and exceeds the Schwarzschild boundary at 3813 Mm (where $\mathrm{\nabla_{rad} = \nabla_{ad}}$). Inside the thin outer envelope convection zone, $\nabla > \mathrm{\nabla_{ad}}$, indicating superadiabatic conditions where convection cannot efficiently transport heat, and radiation contributes to energy transport.

Following the methodology of \citet{Herwig2023}, which identifies convective boundaries through extrema in the product of $\mathrm{Skew \times Kurtosis}$, we determine from the right panel of Figure~\ref{fig:conv-boundary-M424} that the core convective boundary for run M424 is located at 1460 Mm, while the thin outer envelope convection zone boundary is positioned at 3733 Mm. The left panel shows that going from surface towards the center, velocities decrease steeply until \unit{3660}{\Mm}, confirming this region is well within the radiative zone and beyond the overshoot region from the Schwarzschild boundary.

\section{Envelope Convection Simulation: M484}\label{app:M484}

The M484 simulation focuses exclusively on the envelope convection zone. This convection region is spatially confined, with inner and outer boundaries located at \unit{3850}{\Mm} and \unit{4100}{\Mm}, respectively, making it susceptible to boundary effects. No heating was applied at the inner boundary during the simulation initialization. The cooling rate at the outer boundary was set equal to the core heating rate used in comparative simulations.

\begin{figure}
\includegraphics[width=0.49\textwidth]{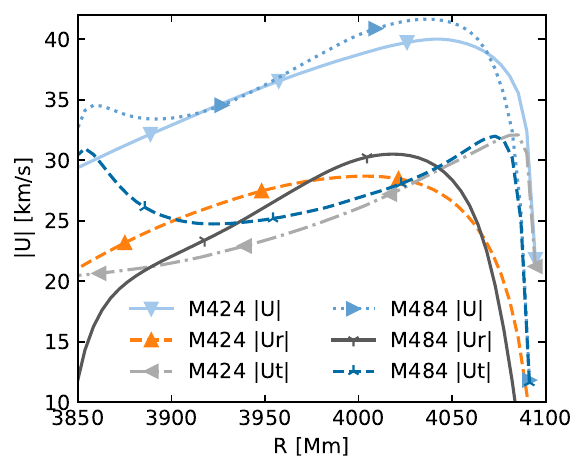}
\includegraphics[width=0.49\textwidth]{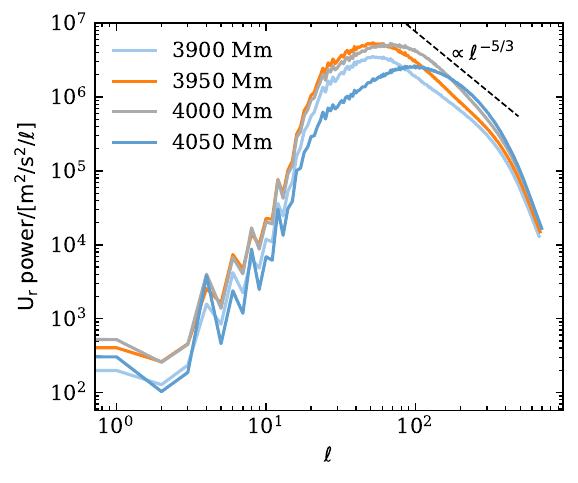}
\caption{(\textit{Left}): Velocity magnitude ($\rm |U|$), radial velocity magnitude ($\rm |Ur|$) and tangential velocity magnitude ($\rm |Ut|$) profiles of the run M424 at dump 4800 and the run M484 at dump 2600 within the envelope convection zone. (\textit{Right}): Radial velocity spatial spectra for simulation M484 at various radial coordinates inside the thin outer envelope convection zone, averaged over last 100 dumps.}
\label{fig:Umag-Spatial_spectra-M484}
\end{figure}

Figure~\ref{fig:Umag-Spatial_spectra-M484} (left panel) shows higher convective velocities in M484 as compared to the full-star run M424. Velocity magnitude differences are evident near the inner boundary because of reflective boundary conditions.

Despite these boundary artifacts, the spatial spectrum (Figure~\ref{fig:Umag-Spatial_spectra-M484}, right panel)) reveals a scale separation between energy injection and dissipation scales. The energy spectrum follows the Kolmogorov power law proportional to $\ell^{-5/3}$, indicating the development of a turbulent cascade within the simulated convection zone.

\section{Convergence}
\begin{figure}
   \centering
   \includegraphics[width=0.49\textwidth]{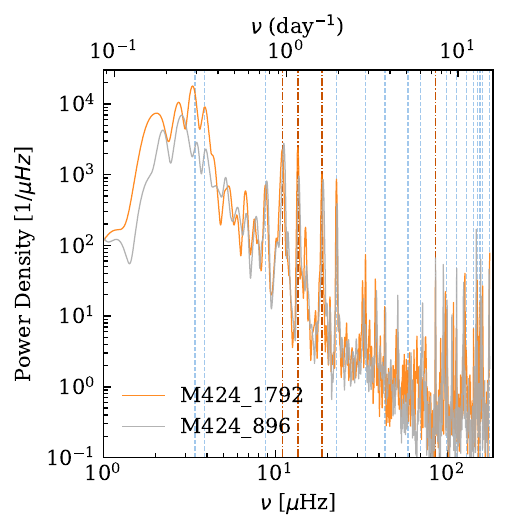}
\includegraphics[width=0.49\textwidth]{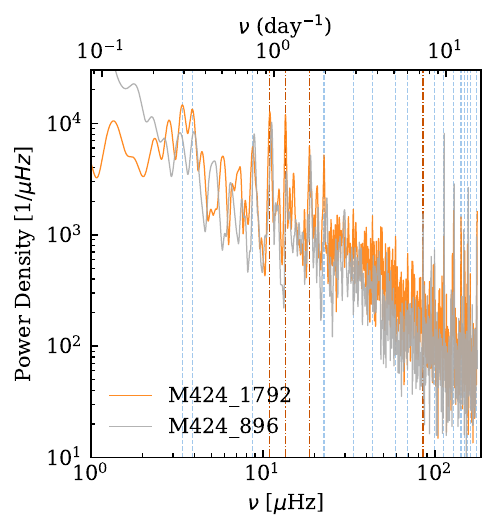}
      \caption{Mock luminosity power spectra $\rm \mathcal{L}_{mock}$ comparison similar to Figure~\ref{Fig:power-spectra-3500Mm} of the full-star run with resolution $1792^3$ presented in the study and with a coarser grid resolution of $896^3$ at \unit{3500}{\Mm} (Left) and \unit{4000}{\Mm} (Right).}
      \label{Fig:Convergence}
\end{figure}
The mock luminosity power spectra of the full-star run with two different grid resolutions $1792^3$ and $896^3$ demonstrate notable convergence characteristics in Figure \ref{Fig:Convergence}. 
\section{Comparison with linear theory}
\begin{figure}
   \centering
   \includegraphics[width=0.98\hsize]{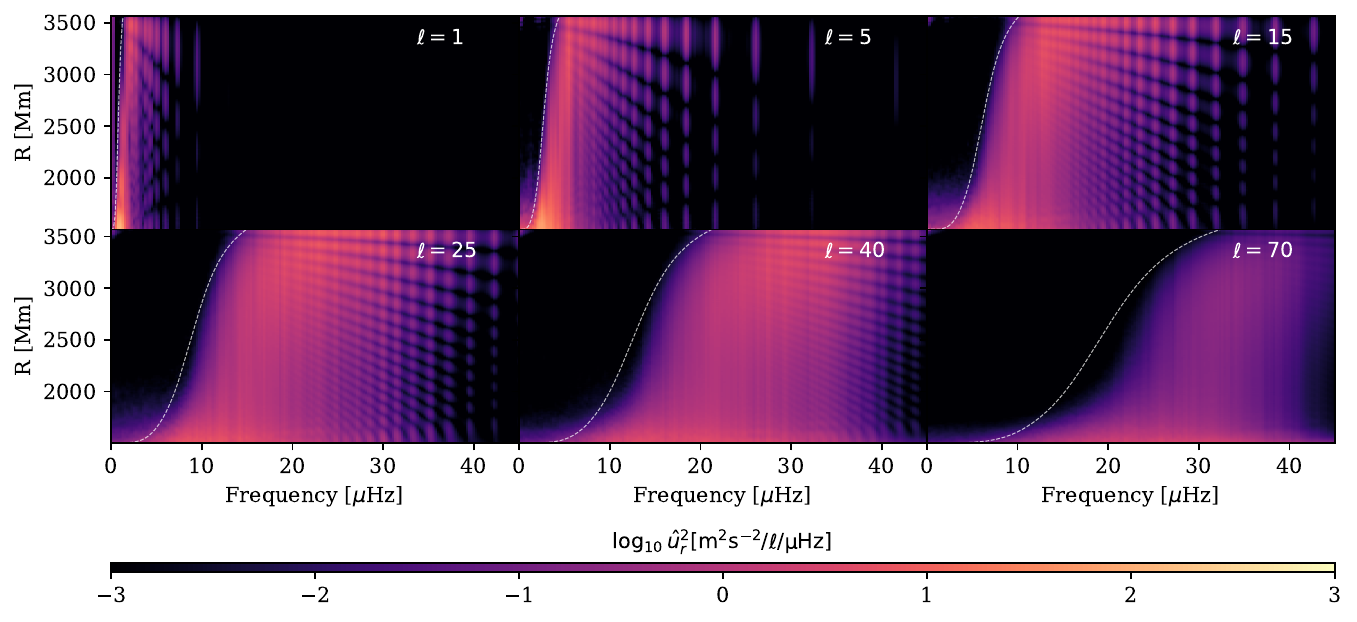}
      \caption{Radial velocity power  $\mathrm{\hat{u}_r^2}$ in different $\ell$ modes
      as a function of radius and frequency for the run M487. The white dashed
      lines indicate the same contour lines as in Figure~\ref{Fig:radiative_damping_with_rho_scaling-M487}.
      }
      \label{Fig:ur_radial_panels-M487}
\end{figure}

\begin{figure}
   \centering
   \includegraphics[width=0.98\hsize]{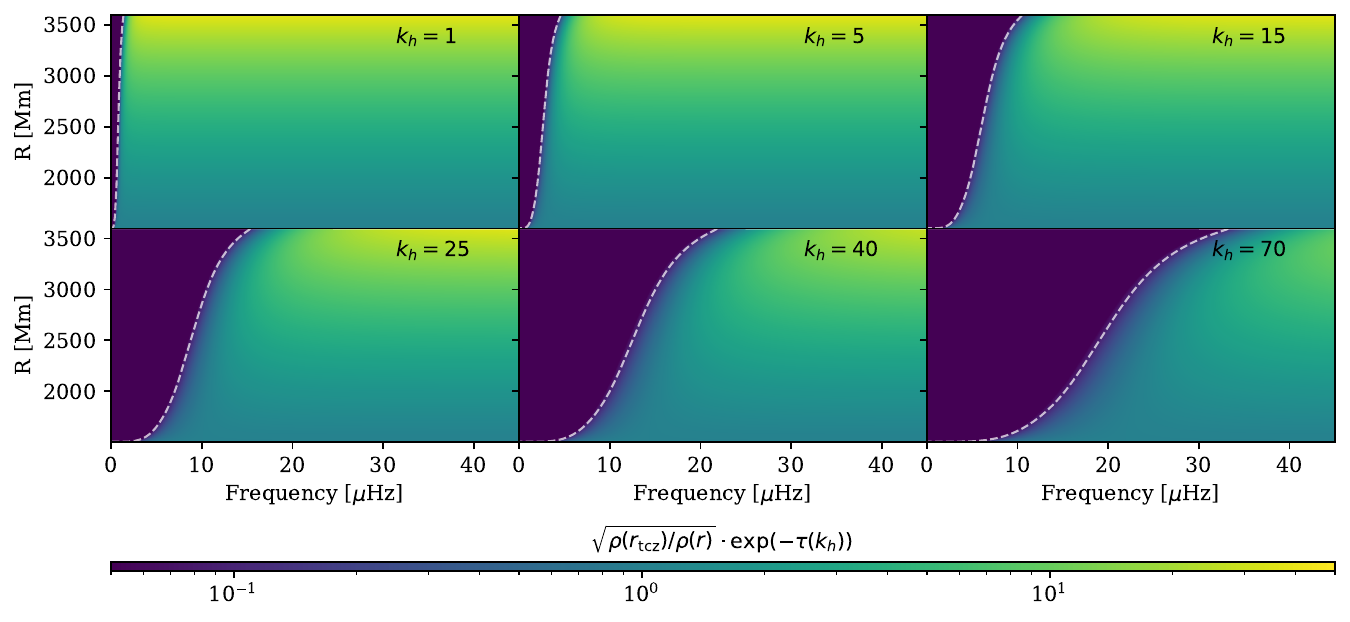}
      \caption{Combined wave amplification and damping factor 
      $\nu_{\mathrm{wave}}(\omega, k_h, r)/\nu_{\mathrm{rms-cz}}$  (cf.\ Equation \ref{eq:wave-damp/amplification}) as a function of radius and frequency for different
      horizontal wavenumbers $k_h$ for the run M487. The white dashed lines indicate the contour 
      where the factor equals $0.1$.
      }
      \label{Fig:radiative_damping_with_rho_scaling-M487}
\end{figure}

The radial velocity power $\mathrm{\hat{u}_r^2}$ decreases at high radii ( Figure~\ref{Fig:ur_radial_panels-M487}). This effect is  confined to low frequencies at low spherical harmonic degrees $\ell$ (top row) and extends to higher frequencies (bottom row) for higher $\ell$ values. The decrease of power with radius is due to radiative damping of IGWs as they propagate outward.  As waves propagate outward in radius, the lowest frequencies experience radiative damping, evidenced by the expansion of the dark region in the upper left portion of each $\ell$ panel. The spatial extent of this damped region shows good agreement with the contour lines derived from linear theory predictions (Equation \ref{eq:wave-damp/amplification}, Figure~\ref{Fig:radiative_damping_with_rho_scaling-M487}). The remaining small differences between the power distribution in the simulation and the radiative damping predictions from linear theory can be attributed to absence of the frequency-dependent IGW driving mechanism, i.e.\ the convective engine, in the prediction shown in Figure~\ref{Fig:radiative_damping_with_rho_scaling-M487}. For example,  for $\ell=1$, substantial power concentrates at the lowest frequencies around 1--3~$\mu$Hz, consistent with the core convection turnover frequency of 1.27~$\mu$Hz.
\end{document}

%% file: macros/units.tex
\newcommand{\numberspace}{\ensuremath{\;}}

\newcommand{\unitstyle}[1]{\ensuremath{\mathrm{#1}}}
\newcommand{\hour}{\unitstyle{h}}

\newcommand{\Mega}{\unitstyle{M}}

\newcommand{\meter}{\unitstyle{m}}

\newcommand{\Kelvin}{\unitstyle{K}}
\newcommand{\K}{\Kelvin}  %degrees Kelvin

\newcommand{\Msun}{\ensuremath{\unitstyle{M}_\odot}}

\newcommand{\Mm}{\Mega\meter}   %megameters
\newcommand{\unit}[2]{\ensuremath{#1\numberspace\mathrm{#2}}}

%% file: references.bib
@ARTICLE{Bowman2023,
       author = {{Bowman}, Dominic M.},
        title = "{Making waves in massive star asteroseismology}",
      journal = {\apss},
         year = 2023,
        month = dec,
       volume = {368},
       number = {12},
          eid = {107},
        pages = {107},
          doi = {10.1007/s10509-023-04262-7},
archivePrefix = {arXiv},
       eprint = {2312.08319},
 primaryClass = {astro-ph.SR},
       adsurl = {https://ui.adsabs.harvard.edu/abs/2023Ap&SS.368..107B}
}

@ARTICLE{Thompson2024,
       author = {{Thompson}, William and {Herwig}, Falk and {Woodward}, Paul R. and {Mao}, Huaqing and {Denissenkov}, Pavel and {Bowman}, Dominic M. and {Blouin}, Simon},
        title = "{3D hydrodynamic simulations of massive main-sequence stars - II. Convective excitation and spectra of internal gravity waves}",
      journal = {\mnras},
         year = 2024,
        month = jun,
       volume = {531},
       number = {1},
        pages = {1316-1337},
          doi = {10.1093/mnras/stae1162},
archivePrefix = {arXiv},
       eprint = {2303.06125},
 primaryClass = {astro-ph.SR},
       adsurl = {https://ui.adsabs.harvard.edu/abs/2024MNRAS.531.1316T}
}

@ARTICLE{Mao2024,
       author = {{Mao}, Huaqing and {Woodward}, Paul and {Herwig}, Falk and {Denissenkov}, Pavel A. and {Blouin}, Simon and {Thompson}, William and {McDermott}, Benjamin},
        title = "{3D Hydrodynamic Simulations of Massive Main-sequence Stars. III. The Effect of Radiation Pressure and Diffusion Leading to a 1D Equilibrium Model}",
      journal = {\apj},
         year = 2024,
        month = nov,
       volume = {975},
       number = {2},
          eid = {271},
        pages = {271},
          doi = {10.3847/1538-4357/ad6c4f},
archivePrefix = {arXiv},
       eprint = {2304.10470},
 primaryClass = {astro-ph.SR},
       adsurl = {https://ui.adsabs.harvard.edu/abs/2024ApJ...975..271M}
}

@ARTICLE{Bowman2019a,
       author = {{Bowman}, D.~M. and {Aerts}, C. and {Johnston}, C. and {Pedersen}, M.~G. and {Rogers}, T.~M. and {Edelmann}, P.~V.~F. and {Sim{\'o}n-D{\'\i}az}, S. and {Van Reeth}, T. and {Buysschaert}, B. and {Tkachenko}, A. and {Triana}, S.~A.},
        title = "{Photometric detection of internal gravity waves in upper main-sequence stars. I. Methodology and application to CoRoT targets}",
      journal = {\aap},
         year = 2019,
        month = jan,
       volume = {621},
          eid = {A135},
        pages = {A135},
          doi = {10.1051/0004-6361/201833662},
archivePrefix = {arXiv},
       eprint = {1811.08023},
 primaryClass = {astro-ph.SR},
       adsurl = {https://ui.adsabs.harvard.edu/abs/2019A&A...621A.135B}
}

@ARTICLE{Bowman2019b,
       author = {{Bowman}, Dominic M. and {Burssens}, Siemen and {Pedersen}, May G. and {Johnston}, Cole and {Aerts}, Conny and {Buysschaert}, Bram and {Michielsen}, Mathias and {Tkachenko}, Andrew and {Rogers}, Tamara M. and {Edelmann}, Philipp V.~F. and {Ratnasingam}, Rathish P. and {Sim{\'o}n-D{\'\i}az}, Sergio and {Castro}, Norberto and {Moravveji}, Ehsan and {Pope}, Benjamin J.~S. and {White}, Timothy R. and {De Cat}, Peter},
        title = "{Low-frequency gravity waves in blue supergiants revealed by high-precision space photometry}",
      journal = {Nature Astronomy},
         year = 2019,
        month = may,
       volume = {3},
        pages = {760-765},
          doi = {10.1038/s41550-019-0768-1},
archivePrefix = {arXiv},
       eprint = {1905.02120},
 primaryClass = {astro-ph.SR},
       adsurl = {https://ui.adsabs.harvard.edu/abs/2019NatAs...3..760B}
}

@ARTICLE{Bowman2024,
       author = {{Bowman}, Dominic M. and {Van Daele}, Pieterjan and {Michielsen}, Mathias and {Van Reeth}, Timothy},
        title = "{Photometric detection of internal gravity waves in upper main-sequence stars: IV. Comparable stochastic low-frequency variability in SMC, LMC, and Galactic massive stars}",
      journal = {\aap},
         year = 2024,
        month = dec,
       volume = {692},
          eid = {A49},
        pages = {A49},
          doi = {10.1051/0004-6361/202451419},
archivePrefix = {arXiv},
       eprint = {2410.12726},
 primaryClass = {astro-ph.SR},
       adsurl = {https://ui.adsabs.harvard.edu/abs/2024A&A...692A..49B}
}

@ARTICLE{Rogers2013,
       author = {{Rogers}, T.~M. and {Lin}, D.~N.~C. and {McElwaine}, J.~N. and {Lau}, H.~H.~B.},
        title = "{Internal Gravity Waves in Massive Stars: Angular Momentum Transport}",
      journal = {\apj},
         year = 2013,
        month = jul,
       volume = {772},
       number = {1},
          eid = {21},
        pages = {21},
          doi = {10.1088/0004-637X/772/1/21},
archivePrefix = {arXiv},
       eprint = {1306.3262},
 primaryClass = {astro-ph.SR},
       adsurl = {https://ui.adsabs.harvard.edu/abs/2013ApJ...772...21R}
}

@ARTICLE{Rogers2015,
       author = {{Rogers}, T.~M.},
        title = "{On the Differential Rotation of Massive Main-sequence Stars}",
      journal = {\apjl},
         year = 2015,
        month = dec,
       volume = {815},
       number = {2},
          eid = {L30},
        pages = {L30},
          doi = {10.1088/2041-8205/815/2/L30},
archivePrefix = {arXiv},
       eprint = {1511.03809},
 primaryClass = {astro-ph.SR},
       adsurl = {https://ui.adsabs.harvard.edu/abs/2015ApJ...815L..30R}
}

@ARTICLE{Rogers2017,
       author = {{Rogers}, T.~M. and {McElwaine}, J.~N.},
        title = "{On the Chemical Mixing Induced by Internal Gravity Waves}",
      journal = {\apjl},
         year = 2017,
        month = oct,
       volume = {848},
       number = {1},
          eid = {L1},
        pages = {L1},
          doi = {10.3847/2041-8213/aa8d13},
archivePrefix = {arXiv},
       eprint = {1709.04920},
 primaryClass = {astro-ph.SR},
       adsurl = {https://ui.adsabs.harvard.edu/abs/2017ApJ...848L...1R}
}

@ARTICLE{Edelmann2019,
       author = {{Edelmann}, P.~V.~F. and {Ratnasingam}, R.~P. and {Pedersen}, M.~G. and {Bowman}, D.~M. and {Prat}, V. and {Rogers}, T.~M.},
        title = "{Three-dimensional Simulations of Massive Stars. I. Wave Generation and Propagation}",
      journal = {\apj},
         year = 2019,
        month = may,
       volume = {876},
       number = {1},
          eid = {4},
        pages = {4},
          doi = {10.3847/1538-4357/ab12df},
archivePrefix = {arXiv},
       eprint = {1903.09392},
 primaryClass = {astro-ph.SR},
       adsurl = {https://ui.adsabs.harvard.edu/abs/2019ApJ...876....4E}
}

@ARTICLE{Ratnasingam2019,
       author = {{Ratnasingam}, R.~P. and {Edelmann}, P.~V.~F. and {Rogers}, T.~M.},
        title = "{Onset of non-linear internal gravity waves in intermediate-mass stars}",
      journal = {\mnras},
         year = 2019,
        month = feb,
       volume = {482},
       number = {4},
        pages = {5500-5512},
          doi = {10.1093/mnras/sty3086},
archivePrefix = {arXiv},
       eprint = {1812.01046},
 primaryClass = {astro-ph.SR},
       adsurl = {https://ui.adsabs.harvard.edu/abs/2019MNRAS.482.5500R}
}

@ARTICLE{Ratnasingam2020,
       author = {{Ratnasingam}, R.~P. and {Edelmann}, P.~V.~F. and {Rogers}, T.~M.},
        title = "{Two-dimensional simulations of internal gravity waves in the radiation zones of intermediate-mass stars}",
      journal = {\mnras},
         year = 2020,
        month = oct,
       volume = {497},
       number = {4},
        pages = {4231-4245},
          doi = {10.1093/mnras/staa2296},
archivePrefix = {arXiv},
       eprint = {2008.03306},
 primaryClass = {astro-ph.SR},
       adsurl = {https://ui.adsabs.harvard.edu/abs/2020MNRAS.497.4231R}
}

@ARTICLE{Ratnasingam2023,
       author = {{Ratnasingam}, R.~P. and {Rogers}, T.~M. and {Chowdhury}, S. and {Handler}, G. and {Vanon}, R. and {Varghese}, A. and {Edelmann}, P.~V.~F.},
        title = "{Internal gravity waves in massive stars. II. Frequency analysis across stellar mass}",
      journal = {\aap},
         year = 2023,
        month = jun,
       volume = {674},
          eid = {A134},
        pages = {A134},
          doi = {10.1051/0004-6361/202245727},
archivePrefix = {arXiv},
       eprint = {2305.06379},
 primaryClass = {astro-ph.SR},
       adsurl = {https://ui.adsabs.harvard.edu/abs/2023A&A...674A.134R}
}

@ARTICLE{Horst2020,
       author = {{Horst}, L. and {Edelmann}, P.~V.~F. and {Andr{\'a}ssy}, R. and {R{\"o}pke}, F.~K. and {Bowman}, D.~M. and {Aerts}, C. and {Ratnasingam}, R.~P.},
        title = "{Fully compressible simulations of waves and core convection in main-sequence stars}",
      journal = {\aap},
         year = 2020,
        month = sep,
       volume = {641},
          eid = {A18},
        pages = {A18},
          doi = {10.1051/0004-6361/202037531},
archivePrefix = {arXiv},
       eprint = {2006.03011},
 primaryClass = {astro-ph.SR},
       adsurl = {https://ui.adsabs.harvard.edu/abs/2020A&A...641A..18H}
}

@ARTICLE{Varghese2023,
       author = {{Varghese}, A. and {Ratnasingam}, R.~P. and {Vanon}, R. and {Edelmann}, P.~V.~F. and {Rogers}, T.~M.},
        title = "{Chemical Mixing Induced by Internal Gravity Waves in Intermediate-mass Stars}",
      journal = {\apj},
         year = 2023,
        month = jan,
       volume = {942},
       number = {1},
          eid = {53},
        pages = {53},
          doi = {10.3847/1538-4357/aca092},
archivePrefix = {arXiv},
       eprint = {2211.06432},
 primaryClass = {astro-ph.SR},
       adsurl = {https://ui.adsabs.harvard.edu/abs/2023ApJ...942...53V}
}

@ARTICLE{Vanon2023,
       author = {{Vanon}, R. and {Edelmann}, P.~V.~F. and {Ratnasingam}, R.~P. and {Varghese}, A. and {Rogers}, T.~M.},
        title = "{Three-dimensional Simulations of Massive Stars. II. Age Dependence}",
      journal = {\apj},
         year = 2023,
        month = sep,
       volume = {954},
       number = {2},
          eid = {171},
        pages = {171},
          doi = {10.3847/1538-4357/ace9db},
archivePrefix = {arXiv},
       eprint = {2307.15109},
 primaryClass = {astro-ph.SR},
       adsurl = {https://ui.adsabs.harvard.edu/abs/2023ApJ...954..171V}
}

@ARTICLE{Anders2023,
       author = {{Anders}, Evan H. and {Lecoanet}, Daniel and {Cantiello}, Matteo and {Burns}, Keaton J. and {Hyatt}, Benjamin A. and {Kaufman}, Emma and {Townsend}, Richard H.~D. and {Brown}, Benjamin P. and {Vasil}, Geoffrey M. and {Oishi}, Jeffrey S. and {Jermyn}, Adam S.},
        title = "{The photometric variability of massive stars due to gravity waves excited by core convection.}",
      journal = {Nature Astronomy},
         year = 2023,
        month = oct,
       volume = {7},
        pages = {1228-1234},
          doi = {10.1038/s41550-023-02040-7},
archivePrefix = {arXiv},
       eprint = {2306.08023},
 primaryClass = {astro-ph.SR},
       adsurl = {https://ui.adsabs.harvard.edu/abs/2023NatAs...7.1228A}
}

@ARTICLE{Lecoanet2019,
       author = {{Lecoanet}, Daniel and {Cantiello}, Matteo and {Quataert}, Eliot and {Couston}, Louis-Alexandre and {Burns}, Keaton J. and {Pope}, Benjamin J.~S. and {Jermyn}, Adam S. and {Favier}, Benjamin and {Le Bars}, Michael},
        title = "{Low-frequency Variability in Massive Stars: Core Generation or Surface Phenomenon?}",
      journal = {\apjl},
         year = 2019,
        month = nov,
       volume = {886},
       number = {1},
          eid = {L15},
        pages = {L15},
          doi = {10.3847/2041-8213/ab5446},
archivePrefix = {arXiv},
       eprint = {1910.01643},
 primaryClass = {astro-ph.SR},
       adsurl = {https://ui.adsabs.harvard.edu/abs/2019ApJ...886L..15L}
}

@ARTICLE{Lecoanet2021,
       author = {{Lecoanet}, Daniel and {Cantiello}, Matteo and {Anders}, Evan H. and {Quataert}, Eliot and {Couston}, Louis-Alexandre and {Bouffard}, Mathieu and {Favier}, Benjamin and {Le Bars}, Michael},
        title = "{Surface manifestation of stochastically excited internal gravity waves}",
      journal = {\mnras},
         year = 2021,
        month = nov,
       volume = {508},
       number = {1},
        pages = {132-143},
          doi = {10.1093/mnras/stab2524},
archivePrefix = {arXiv},
       eprint = {2105.04558},
 primaryClass = {astro-ph.SR},
       adsurl = {https://ui.adsabs.harvard.edu/abs/2021MNRAS.508..132L}
}

@ARTICLE{LeSaux2023,
       author = {{Le Saux}, A. and {Baraffe}, I. and {Guillet}, T. and {Vlaykov}, D.~G. and {Morison}, A. and {Pratt}, J. and {Constantino}, T. and {Goffrey}, T.},
        title = "{Two-dimensional simulations of internal gravity waves in a 5 M$_{{\ensuremath{\odot}}}$ zero-age-main-sequence model}",
      journal = {\mnras},
         year = 2023,
        month = jun,
       volume = {522},
       number = {2},
        pages = {2835-2849},
          doi = {10.1093/mnras/stad1067},
archivePrefix = {arXiv},
       eprint = {2304.02508},
 primaryClass = {astro-ph.SR},
       adsurl = {https://ui.adsabs.harvard.edu/abs/2023MNRAS.522.2835L}
}

@ARTICLE{Cantiello2009,
       author = {{Cantiello}, M. and {Langer}, N. and {Brott}, I. and {de Koter}, A. and {Shore}, S.~N. and {Vink}, J.~S. and {Voegler}, A. and {Lennon}, D.~J. and {Yoon}, S. -C.},
        title = "{Sub-surface convection zones in hot massive stars and their observable consequences}",
      journal = {\aap},
         year = 2009,
        month = may,
       volume = {499},
       number = {1},
        pages = {279-290},
          doi = {10.1051/0004-6361/200911643},
archivePrefix = {arXiv},
       eprint = {0903.2049},
 primaryClass = {astro-ph.SR},
       adsurl = {https://ui.adsabs.harvard.edu/abs/2009A&A...499..279C}
}

@ARTICLE{Cantiello2021,
       author = {{Cantiello}, Matteo and {Lecoanet}, Daniel and {Jermyn}, Adam S. and {Grassitelli}, Luca},
        title = "{On the Origin of Stochastic, Low-Frequency Photometric Variability in Massive Stars}",
      journal = {\apj},
         year = 2021,
        month = jul,
       volume = {915},
       number = {2},
          eid = {112},
        pages = {112},
          doi = {10.3847/1538-4357/ac03b0},
archivePrefix = {arXiv},
       eprint = {2102.05670},
 primaryClass = {astro-ph.SR},
       adsurl = {https://ui.adsabs.harvard.edu/abs/2021ApJ...915..112C}
}

@ARTICLE{Schultz2022,
       author = {{Schultz}, William C. and {Bildsten}, Lars and {Jiang}, Yan-Fei},
        title = "{Stochastic Low-frequency Variability in Three-dimensional Radiation Hydrodynamical Models of Massive Star Envelopes}",
      journal = {\apjl},
         year = 2022,
        month = jan,
       volume = {924},
       number = {1},
          eid = {L11},
        pages = {L11},
          doi = {10.3847/2041-8213/ac441f},
archivePrefix = {arXiv},
       eprint = {2110.13944},
 primaryClass = {astro-ph.SR},
       adsurl = {https://ui.adsabs.harvard.edu/abs/2022ApJ...924L..11S}
}

@ARTICLE{Jermyn2022,
       author = {{Jermyn}, Adam S. and {Anders}, Evan H. and {Cantiello}, Matteo},
        title = "{A Transparent Window into Early-type Stellar Variability}",
      journal = {\apj},
         year = 2022,
        month = feb,
       volume = {926},
       number = {2},
          eid = {221},
        pages = {221},
          doi = {10.3847/1538-4357/ac4e89},
archivePrefix = {arXiv},
       eprint = {2201.10567},
 primaryClass = {astro-ph.SR},
       adsurl = {https://ui.adsabs.harvard.edu/abs/2022ApJ...926..221J}
}

@ARTICLE{Mombarg2024,
       author = {{Mombarg}, Joey S.~G. and {Aerts}, Conny and {Van Reeth}, Timothy and {Hey}, Daniel},
        title = "{Estimates of (convective core) masses, radii, and relative ages for {\ensuremath{\sim}}14 000 Gaia-discovered gravity-mode pulsators monitored by TESS}",
      journal = {\aap},
         year = 2024,
        month = nov,
       volume = {691},
          eid = {A131},
        pages = {A131},
          doi = {10.1051/0004-6361/202451651},
archivePrefix = {arXiv},
       eprint = {2410.05367},
 primaryClass = {astro-ph.SR},
       adsurl = {https://ui.adsabs.harvard.edu/abs/2024A&A...691A.131M}
}

@ARTICLE{Aerts2021,
       author = {{Aerts}, C.},
        title = "{Probing the interior physics of stars through asteroseismology}",
      journal = {Reviews of Modern Physics},
         year = 2021,
        month = jan,
       volume = {93},
       number = {1},
          eid = {015001},
        pages = {015001},
          doi = {10.1103/RevModPhys.93.015001},
archivePrefix = {arXiv},
       eprint = {1912.12300},
 primaryClass = {astro-ph.SR},
       adsurl = {https://ui.adsabs.harvard.edu/abs/2021RvMP...93a5001A}
}

@ARTICLE{Woodward2015,
       author = {{Woodward}, Paul R. and {Herwig}, Falk and {Lin}, Pei-Hung},
        title = "{Hydrodynamic Simulations of H Entrainment at the Top of He-shell Flash Convection}",
      journal = {\apj},
         year = 2015,
        month = jan,
       volume = {798},
       number = {1},
          eid = {49},
        pages = {49},
          doi = {10.1088/0004-637X/798/1/49},
       adsurl = {https://ui.adsabs.harvard.edu/abs/2015ApJ...798...49W}
}

@ARTICLE{Herwig2023,
       author = {{Herwig}, Falk and {Woodward}, Paul R. and {Mao}, Huaqing and {Thompson}, William R. and {Denissenkov}, Pavel and {Lau}, Josh and {Blouin}, Simon and {Andrassy}, Robert and {Paul}, Adam},
        title = "{3D hydrodynamic simulations of massive main-sequence stars - I. Dynamics and mixing of convection and internal gravity waves}",
      journal = {\mnras},
         year = 2023,
        month = oct,
       volume = {525},
       number = {2},
        pages = {1601-1629},
          doi = {10.1093/mnras/stad2157},
archivePrefix = {arXiv},
       eprint = {2303.05495},
 primaryClass = {astro-ph.SR},
       adsurl = {https://ui.adsabs.harvard.edu/abs/2023MNRAS.525.1601H}
}

@ARTICLE{Colella1984,
       author = {{Colella}, P. and {Woodward}, Paul R.},
        title = "{The Piecewise Parabolic Method (PPM) for Gas-Dynamical Simulations}",
      journal = {Journal of Computational Physics},
         year = 1984,
        month = sep,
       volume = {54},
        pages = {174-201},
          doi = {10.1016/0021-9991(84)90143-8},
       adsurl = {https://ui.adsabs.harvard.edu/abs/1984JCoPh..54..174C}
}

@BOOK{Aerts2010,
       author = {{Aerts}, Conny and {Christensen-Dalsgaard}, J{\o}rgen and {Kurtz}, Donald W.},
        title = "{Asteroseismology}",
         year = 2010,
          doi = {10.1007/978-1-4020-5803-5},
       adsurl = {https://ui.adsabs.harvard.edu/abs/2010aste.book.....A}
}

@ARTICLE{Paxton2011,
       author = {{Paxton}, Bill and {Bildsten}, Lars and {Dotter}, Aaron and {Herwig}, Falk and {Lesaffre}, Pierre and {Timmes}, Frank},
        title = "{Modules for Experiments in Stellar Astrophysics (MESA)}",
      journal = {\apjs},
         year = 2011,
        month = jan,
       volume = {192},
       number = {1},
          eid = {3},
        pages = {3},
          doi = {10.1088/0067-0049/192/1/3},
archivePrefix = {arXiv},
       eprint = {1009.1622},
 primaryClass = {astro-ph.SR},
       adsurl = {https://ui.adsabs.harvard.edu/abs/2011ApJS..192....3P}
}

@ARTICLE{Aerts2015,
       author = {{Aerts}, C. and {Rogers}, T.~M.},
        title = "{Observational Signatures of Convectively Driven Waves in Massive Stars}",
      journal = {\apjl},
         year = 2015,
        month = jun,
       volume = {806},
       number = {2},
          eid = {L33},
        pages = {L33},
          doi = {10.1088/2041-8205/806/2/L33},
archivePrefix = {arXiv},
       eprint = {1505.06648},
 primaryClass = {astro-ph.SR},
       adsurl = {https://ui.adsabs.harvard.edu/abs/2015ApJ...806L..33A}
}

@ARTICLE{Blomme2011,
       author = {{Blomme}, R. and {Mahy}, L. and {Catala}, C. and {Cuypers}, J. and {Gosset}, E. and {Godart}, M. and {Montalban}, J. and {Ventura}, P. and {Rauw}, G. and {Morel}, T. and {Degroote}, P. and {Aerts}, C. and {Noels}, A. and {Michel}, E. and {Baudin}, F. and {Baglin}, A. and {Auvergne}, M. and {Samadi}, R.},
        title = "{Variability in the CoRoT photometry of three hot O-type stars. HD 46223, HD 46150, and HD 46966}",
      journal = {\aap},
         year = 2011,
        month = sep,
       volume = {533},
          eid = {A4},
        pages = {A4},
          doi = {10.1051/0004-6361/201116949},
archivePrefix = {arXiv},
       eprint = {1107.0267},
 primaryClass = {astro-ph.SR},
       adsurl = {https://ui.adsabs.harvard.edu/abs/2011A&A...533A...4B}
}

@ARTICLE{Rider1999,
       author = {{Rider}, William J. and {Knoll}, Dana A.},
        title = "{Time Step Size Selection for Radiation Diffusion Calculations}",
      journal = {Journal of Computational Physics},
         year = 1999,
        month = jul,
       volume = {152},
       number = {2},
        pages = {790-795},
          doi = {10.1006/jcph.1999.6266},
       adsurl = {https://ui.adsabs.harvard.edu/abs/1999JCoPh.152..790R}
}

@ARTICLE{Courant1928,
       author = {{Courant}, R. and {Friedrichs}, K. and {Lewy}, H.},
        title = "{{\"U}ber die partiellen Differenzengleichungen der mathematischen Physik}",
      journal = {Mathematische Annalen},
         year = 1928,
        month = jan,
       volume = {100},
        pages = {32-74},
          doi = {10.1007/BF01448839},
       adsurl = {https://ui.adsabs.harvard.edu/abs/1928MatAn.100...32C}
}

@ARTICLE{Kumar1999,
       author = {{Kumar}, Pawan and {Talon}, Suzanne and {Zahn}, Jean-Paul},
        title = "{Angular Momentum Redistribution by Waves in the Sun}",
      journal = {\apj},
         year = 1999,
        month = aug,
       volume = {520},
       number = {2},
        pages = {859-870},
          doi = {10.1086/307464},
archivePrefix = {arXiv},
       eprint = {astro-ph/9902309},
 primaryClass = {astro-ph},
       adsurl = {https://ui.adsabs.harvard.edu/abs/1999ApJ...520..859K}
}

@ARTICLE{Krticka2021,
       author = {{Krti{\v{c}}ka}, J. and {Feldmeier}, A.},
        title = "{Stochastic light variations in hot stars from wind instability: finding photometric signatures and testing against the TESS data}",
      journal = {\aap},
         year = 2021,
        month = apr,
       volume = {648},
          eid = {A79},
        pages = {A79},
          doi = {10.1051/0004-6361/202040148},
archivePrefix = {arXiv},
       eprint = {2103.08755},
 primaryClass = {astro-ph.SR},
       adsurl = {https://ui.adsabs.harvard.edu/abs/2021A&A...648A..79K}
}

@ARTICLE{Krticka2018,
       author = {{Krti{\v{c}}ka}, J. and {Feldmeier}, A.},
        title = "{Light variations due to the line-driven wind instability and wind blanketing in O stars}",
      journal = {\aap},
         year = 2018,
        month = sep,
       volume = {617},
          eid = {A121},
        pages = {A121},
          doi = {10.1051/0004-6361/201731614},
archivePrefix = {arXiv},
       eprint = {1807.09407},
 primaryClass = {astro-ph.SR},
       adsurl = {https://ui.adsabs.harvard.edu/abs/2018A&A...617A.121K}
}

@dataset{Bowman2020,
       author = {{Bowman}, D.~M. and {Burssens}, S. and {Simon-Diaz}, S. and {Edelmann}, P.~V.~F. and {Rogers}, T.~M. and {Horst}, L. and {Ropke}, F.~K. and {Aerts}, C.},
        title = "{VizieR Online Data Catalog: OB stars TESS phot. \& high-resolution spectroscopy (Bowman+, 2020)}",
 howpublished = {VizieR On-line Data Catalog: J/A+A/640/A36. Originally published in: 2020A\&A...640A..36B},
         year = 2020,
        month = oct,
          eid = {J/A+A/640/A36},
          doi = {10.26093/cds/vizier.36400036},
       adsurl = {https://ui.adsabs.harvard.edu/abs/2020yCat..36400036B}
}

@ARTICLE{Schultz2023,
       author = {{Schultz}, William C. and {Bildsten}, Lars and {Jiang}, Yan-Fei},
        title = "{Turbulence-supported Massive Star Envelopes}",
      journal = {\apjl},
         year = 2023,
        month = jul,
       volume = {951},
       number = {2},
          eid = {L42},
        pages = {L42},
          doi = {10.3847/2041-8213/acdf50},
archivePrefix = {arXiv},
       eprint = {2306.08034},
 primaryClass = {astro-ph.SR},
       adsurl = {https://ui.adsabs.harvard.edu/abs/2023ApJ...951L..42S}
}

@ARTICLE{Stephens2021,
       author = {{Stephens}, David and {Herwig}, Falk and {Woodward}, Paul and {Denissenkov}, Pavel and {Andrassy}, Robert and {Mao}, Huaqing},
        title = "{3D1D hydro-nucleosynthesis simulations - I. Advective-reactive post-processing method and its application to H ingestion into He-shell flash convection in rapidly accreting white dwarfs}",
      journal = {\mnras},
         year = 2021,
        month = jun,
       volume = {504},
       number = {1},
        pages = {744-760},
          doi = {10.1093/mnras/stab500},
archivePrefix = {arXiv},
       eprint = {2001.10969},
 primaryClass = {astro-ph.SR},
       adsurl = {https://ui.adsabs.harvard.edu/abs/2021MNRAS.504..744S}
}

@ARTICLE{Jones2017,
       author = {{Jones}, S. and {Andrassy}, R. and {Sandalski}, S. and {Davis}, A. and {Woodward}, P. and {Herwig}, F.},
        title = "{Idealized hydrodynamic simulations of turbulent oxygen-burning shell convection in 4{\ensuremath{\pi}} geometry}",
      journal = {\mnras},
         year = 2017,
        month = mar,
       volume = {465},
       number = {3},
        pages = {2991-3010},
          doi = {10.1093/mnras/stw2783},
archivePrefix = {arXiv},
       eprint = {1605.03766},
 primaryClass = {astro-ph.SR},
       adsurl = {https://ui.adsabs.harvard.edu/abs/2017MNRAS.465.2991J}
}

@book{Buehler2009, place={Cambridge}, series={Cambridge Monographs on Mechanics}, title={Waves and Mean Flows}, publisher={Cambridge University Press}, author={B\"uhler, Oliver}, year={2009}, collection={Cambridge Monographs on Mechanics}}

@ARTICLE{Burssens2023,
       author = {{Burssens}, Siemen and {Bowman}, Dominic M. and {Michielsen}, Mathias and {Sim{\'o}n-D{\'\i}az}, Sergio and {Aerts}, Conny and {Vanlaer}, Vincent and {Banyard}, Gareth and {Nardetto}, Nicolas and {Townsend}, Richard H.~D. and {Handler}, Gerald and {Mombarg}, Joey S.~G. and {Vanderspek}, Roland and {Ricker}, George},
        title = "{A calibration point for stellar evolution from massive star asteroseismology}",
      journal = {Nature Astronomy},
         year = 2023,
        month = aug,
       volume = {7},
        pages = {913-930},
          doi = {10.1038/s41550-023-01978-y},
archivePrefix = {arXiv},
       eprint = {2306.11798},
 primaryClass = {astro-ph.SR},
       adsurl = {https://ui.adsabs.harvard.edu/abs/2023NatAs...7..913B}
}

@ARTICLE{Pedersen2018,
       author = {{Pedersen}, M.~G. and {Aerts}, C. and {P{\'a}pics}, P.~I. and {Rogers}, T.~M.},
        title = "{The shape of convective core overshooting from gravity-mode period spacings}",
      journal = {\aap},
         year = 2018,
        month = jul,
       volume = {614},
          eid = {A128},
        pages = {A128},
          doi = {10.1051/0004-6361/201732317},
archivePrefix = {arXiv},
       eprint = {1802.02051},
 primaryClass = {astro-ph.SR},
       adsurl = {https://ui.adsabs.harvard.edu/abs/2018A&A...614A.128P}
}

@ARTICLE{Papics2017,
       author = {{P{\'a}pics}, P.~I. and {Tkachenko}, A. and {Van Reeth}, T. and {Aerts}, C. and {Moravveji}, E. and {Van de Sande}, M. and {De Smedt}, K. and {Bloemen}, S. and {Southworth}, J. and {Debosscher}, J. and {Niemczura}, E. and {Gameiro}, J.~F.},
        title = "{Signatures of internal rotation discovered in the Kepler data of five slowly pulsating B stars}",
      journal = {\aap},
         year = 2017,
        month = feb,
       volume = {598},
          eid = {A74},
        pages = {A74},
          doi = {10.1051/0004-6361/201629814},
archivePrefix = {arXiv},
       eprint = {1611.06955},
 primaryClass = {astro-ph.SR},
       adsurl = {https://ui.adsabs.harvard.edu/abs/2017A&A...598A..74P}
}

@ARTICLE{Woodward1984,
       author = {{Woodward}, P. and {Colella}, P.},
        title = "{The Numerical Stimulation of Two-Dimensional Fluid Flow with Strong Shocks}",
      journal = {Journal of Computational Physics},
         year = 1984,
        month = apr,
       volume = {54},
       number = {1},
        pages = {115-173},
          doi = {10.1016/0021-9991(84)90142-6},
       adsurl = {https://ui.adsabs.harvard.edu/abs/1984JCoPh..54..115W}
}

@ARTICLE{Porter2000,
       author = {{Porter}, David H. and {Woodward}, Paul R.},
        title = "{Three-dimensional Simulations of Turbulent Compressible Convection}",
      journal = {\apjs},
         year = 2000,
        month = mar,
       volume = {127},
       number = {1},
        pages = {159-187},
          doi = {10.1086/313310},
       adsurl = {https://ui.adsabs.harvard.edu/abs/2000ApJS..127..159P}
}

@ARTICLE{Zahn1997,
       author = {{Zahn}, J. -P. and {Talon}, S. and {Matias}, J.},
        title = "{Angular momentum transport by internal waves in the solar interior.}",
      journal = {\aap},
         year = 1997,
        month = jun,
       volume = {322},
        pages = {320-328},
          doi = {10.48550/arXiv.astro-ph/9611189},
archivePrefix = {arXiv},
       eprint = {astro-ph/9611189},
 primaryClass = {astro-ph},
       adsurl = {https://ui.adsabs.harvard.edu/abs/1997A&A...322..320Z}
}

@ARTICLE{Blouin2023,
       author = {{Blouin}, Simon and {Mao}, Huaqing and {Herwig}, Falk and {Denissenkov}, Pavel and {Woodward}, Paul R. and {Thompson}, William R.},
        title = "{3D hydrodynamics simulations of internal gravity waves in red giant branch stars}",
      journal = {\mnras},
         year = 2023,
        month = jun,
       volume = {522},
       number = {2},
        pages = {1706-1725},
          doi = {10.1093/mnras/stad1115},
archivePrefix = {arXiv},
       eprint = {2303.07332},
 primaryClass = {astro-ph.SR},
       adsurl = {https://ui.adsabs.harvard.edu/abs/2023MNRAS.522.1706B}
}

@ARTICLE{Pedersen2025,
       author = {{Pedersen}, May G. and {Bildsten}, Lars},
        title = "{Stochastic low-frequency variability of 50 massive stars in the Cygnus OB associations and the Small Magellanic Cloud}",
      journal = {\mnras},
         year = 2025,
        month = may,
       volume = {539},
       number = {3},
        pages = {2742-2764},
          doi = {10.1093/mnras/staf661},
archivePrefix = {arXiv},
       eprint = {2504.15861},
 primaryClass = {astro-ph.SR},
       adsurl = {https://ui.adsabs.harvard.edu/abs/2025MNRAS.539.2742P}
}
